\documentclass[useAMS,usenatbib,usegraphicx]{mn2e}

\usepackage{amsmath}
\usepackage{amsfonts}
\usepackage{rotating}
\usepackage{color}

\newcommand{\aap}{A\&A} 
\newcommand{\aaps}{A\&AS} 
\newcommand{\mnras}{MNRAS} 
\newcommand{\apj}{ApJ} 
\newcommand{\araa}{ARA\&A} 

\newcommand{\apjs}{ApJS}
\newcommand{\aj}{AJ}
\newcommand{\pasp}{PASP} 
\newcommand{\arep}{Astron. Rep.}
\newcommand{\rmp}{Rev. Modern Phys.}

\newcommand{\teff}{$T_{\mbox{\scriptsize eff}}$}
\newcommand{\kms}{$\mbox{km\,s}^{-1}$}

\title[Elemental abundances of the open cluster IC 4756]{High-resolution elemental abundance analysis of the open cluster IC 4756}
\author[Y.-S.~Ting et al.]{Yuan-Sen~Ting,$^{1,2}$\thanks{E-mail: yuan-sen.ting@cfa.harvard.com} Gayandhi~M.~De~Silva,$^{3}$ Kenneth~C.~Freeman$^{1}$ and 
\newauthor Stacey-Jo~Parker,$^{4}$ \\
$^{1}$Research School of Astronomy $\&$ Astrophysics, The Australian National University, Cotter Road, Weston Creek, ACT 2611, Australia\\
$^{2}$Harvard-Smithsonian Center for Astrophysics, 60 Garden Street, Cambridge, MA 02139, USA\\
$^{3}$Australian Astronomical Observatory, PO Box 296, NSW 1710, Australia \\
$^{4}$Institute of Astronomy, University of Cambridge, Madingley Road, Cambridge CB3 OHA
}
\begin{document}

\date{Accepted 2012 August 31. Received 2012 August 25; in original form 2012 July 30}

\pagerange{0--0} \pubyear{2012}

\maketitle

\label{firstpage}

\begin{abstract}
We present detailed elemental abundances of 12 subgiants in the open cluster IC 4756 including Na, Al, Mg, Si, Ca, Ti, Cr, Ni, Fe, Zn and Ba. We measure the cluster to have [Fe/H] = $-0.01\pm0.10$. Most of the measured star-to-star [X/H] abundance variation is below $\sigma < 0.03$, as expected from a coeval stellar population preserving natal abundance patterns, supporting the use of elemental abundances as a probe to reconstruct dispersed clusters.  We find discrepancies between {Cr\,{\sc i}} and {Cr\,{\sc ii}} abundances as well as between {Ti\,{\sc i}} and {Ti\,{\sc ii}} abundances, where the ionized abundances are larger by about 0.2~dex. This follows other such studies which demonstrate the effects of overionization in cool stars. IC 4756 are supersolar in Mg, Si, Na and Al, but are solar in the other elements. The fact that IC 4756 is supersolar in some $\alpha$-elements (Mg, Si) but solar in the others (Ca, Ti) suggests that the production of $\alpha$-elements is not simply one dimensional and could be exploited for chemical tagging.
\end{abstract}
 
\begin{keywords}
stars: abundances -- ISM: abundances -- ISM: evolution.
\end{keywords}

\section[]{Introduction}
\label{section:introduction}

Understanding the formation of the Milky Way is a very challenging task. Among the major challenges are the following: (1) although stars form in aggregates, they disperse due to dynamical influences, and the kinematical data we can measure today, such as radial velocity and proper motions, carry little information on the site and epoch of the star formation and (2) we are restricted to a small survey volume around the solar neighborhood.

Measuring elemental abundances of stars in open clusters/moving groups with high-resolution spectrograph helps us to shed light on these problems. Elemental abundances are shown to be powerful tools to uncover the natal conditions of stars \citep*{free02} because stars that were formed with the same gas cloud should carry the same pollution history of the gas, either due to supernova ejecta \citep*[e.g][]{woo95,chi02,kob06} or due to stellar mass loss from low-mass stars especially during the asymptotic giant branch (AGB) phase \citep*[e.g.][]{vas93,kar10}. A growing number of recent studies \citep*[e.g.][Ting et al. in preparation]{bla10,bov12a,bov12b,bov12c} show that chemical tagging, namely reconstructing ancient clusters using elemental abundances, is a promising approach to understand Galactic dynamics and Milky Way substructures. 

Most of the field star spectroscopic surveys carried out to date cover a small volume around the solar neighbourhood with $7.5 \la R_G \la 8.5$ kpc, where $R_G$ is the distance from the Galactic Centre. Obtaining high-resolution spectra of field stars at larger Galactic radius proved to be observationally challenging. Fortunately, open clusters and moving groups provide us with unique opportunity to probe larger Galactic radius where chemical homogeneity within such substructures have been established, at least for heavier $(Z \geq 12)$ elements \citep*[e.g.][]{siv07b,siv11,bub10,fri10,pan10,jac11}. 

In this paper, we target the open cluster IC 4756.  Located 484~pc away, it is an intermediate-aged open cluster similar to the Hyades with an age of about 790 Myr \citep*{sal04}. The metallicity has not been well constrained in the literature with a range from [Fe/H]\footnote{By definition, $\mbox{[X/Y]} \equiv \log_{10} (N_{\mbox{\scriptsize X}}/N_{\mbox{\scriptsize Y}})_{\mbox{star}} - \log_{10} (N_{\mbox{\scriptsize X}}/N_{\mbox{\scriptsize Y}})_{\odot}$, where $N_{\mbox{\scriptsize X}}$ and $N_{\mbox{\scriptsize Y}}$ are the abundances of element X and element Y, respectively. } $\simeq$ -0.22 \citep*{tho93} to more recent values of 0.03 \citep{san09} and 0.05 \citep{smi09} using high-resolution spectroscopy. IC 4756 was previously studied by various groups, including \citet{gil89}, \citet{luc94}, \citet{jac07}, \citet{san09}, \citet{smi09} and \citet{pac10}. Most recent authors, including this study, found that IC 4756 has about solar metallicity. However, \citet{jac07} found that [Fe/H] $\simeq$ -0.15. 

In this paper, we explore the abundances of 12 subgiants in IC 4756, most of which have previously not been studied, thereby extending the total sample size. Further, we extend the list of elements being measured by including Zn and Ba that have not been measured with a large and homogeneous sample. \citet{ting12} showed that Fe-peak elements such as Zn are important to distinguish different origin of clusters or stars that have distinctive pollution history, presumably from more energetic core-collapse supernovae \citep{ume02,ume05}. They also showed that neutron-capture elements, such as Ba, are crucial to distinguish the origin of metal-rich stars and clusters, as it is a telltale imprint of chemical evolution via the AGB slow-process ($s$-process) nucleosynthesis mechanism \citep*[for a review, see][]{bus99,her05,kap11}. 

Although it is argued that heavier elements' abundances are mostly homogeneous across the same open cluster, with the expectation that these abundances are unchanged during stellar evolution, only about 5 per cent of open clusters have been studied with high-resolution spectroscopy \citep*[e.g.][]{siv06,siv07a,pan10,roe11}. We re-examine the homogeneity of elemental abundances up to Ba, and that is one of the main purposes of this paper. The analysis presented in the paper will complement the upcoming large-scale high-resolution multi-object surveys, such as the Galactic Archaeology with HERMES (GALAH) survey, which assumes homogeneity of elemental abundances to reconstruct ancient clusters.

\section[]{Observations}\label{sec:observation}

\subsection[]{Observations}\label{subsec:observation}

Our sample contains 12 subgiants selected from \citet*{her75} catalogue with proper motions and photometric memberships available at the time of observation. Targets were of G--K spectral type and were selected in order to avoid rapid rotators. The enumeration of the stars are based on \citet{her75} and are listed in Table~\ref{tab:observations}. 

High-resolution spectra for the selected stars were obtained using the echelle spectrograph on the Astrophysical Research Consortium (ARC) 3.5-m telescope at Apache Point Observatory. The Nasmyth-mounted echelle spectrograph provides a resolution of  $R \simeq 30\,000$ and full wavelength coverage from approximately $3\,500$ to $10\,000$\,\AA. The typical signal-to-noise ratio of the final co-added spectra was $>100$ per pixel at $5\,500$\,\AA. The data reduction, including de-biasing, flat-fielding, scattered light removal, order extraction, wavelength calibration and continuum fitting, were carried out using {\sc iraf}\footnote{{\sc iraf} is distributed by the National Optical Astronomy Observatory, which is operated by the Association of Universities for Research in Astronomy, Inc., under cooperative agreement with the National Science Foundation} {\sc echelle} package.  

Our radial velocity measurements and subsequent spectroscopic analysis, as discussed below, show that one star in the sample, Her 85, is a non-member of the cluster. It had a radial velocity of -20.8 \kms, which deviates from the cluster mean velocity as well as low iron abundances ([{Fe\,{\sc i}}/H]$=7.14$ and [{Fe\,{\sc ii}}/H]$=7.06$) from our analysis. Therefore we discard Her 85 from further analysis and discussion.

\begin{table}
\begin{center}
\caption{Stellar Sample.\label{tab:observations}}
\begin{tabular}{lccccc}
\hline 
Star    &$\,$& $V$  & $B-V$ &  RA          &  Dec             \\
\hline
Her 6   &$\,$& 8.97 & 1.26  & 18 36 33.227 & +05 12 42.78     \\
Her 35  &$\,$& 9.66 & 0.98  & 18 37 05.217 & +05 17 31.60     \\
Her 82  &$\,$& 9.78 & 1.02  & 18 37 30.299 & +05 12 15.72     \\
Her 87  &$\,$& 9.43 & 1.16  & 18 37 34.219 & +05 28 33.47     \\
Her 90  &$\,$& 8.00 & 1.36  & 18 37 35.828 & +05 15 37.82     \\
Her 144 &$\,$& 9.23 & 1.02  & 18 38 05.162 & +05 24 33.76     \\
Her 170 &$\,$& 9.48 & 0.97  & 18 38 17.578 & +05 38 17.04     \\
Her 176 &$\,$& 9.42 & 0.97  & 18 38 20.759 & +05 26 02.31     \\
Her 228 &$\,$& 9.41 & 0.98  & 18 38 43.788 & +05 14 19.96     \\    
Her 249 &$\,$& 9.04 & 1.07  & 18 38 52.930 & +05 20 16.52     \\
Her 296 &$\,$& 9.29 & 0.97  & 18 39 17.877 & +05 13 48.78     \\     
Her 397 &$\,$& 9.23 & 1.05  & 18 40 18.515 & +05 18 51.74     \\
\hline
\end{tabular}
\end{center}
\end{table}

\section[]{Abundance Analysis}\label{sec:data}

\begin{table}
\begin{center}
\caption{Spectroscopic determined atmospheric parameters.\label{tab:parameters}}
\begin{tabular}{lccccccc}
\hline 
& \multicolumn{3}{c}{This study} & \multicolumn{3}{c}{\citet{gil89}}\\[0.05cm]
\\[-0.25cm]
        & \teff & $\log g$ & $v_t$  & \teff & $\log g$ & $v_t$   \\
Star    & (K)   & (dex)    & (\kms) &  (K)  & (dex)    & (\kms)  \\
\hline
Her 6   &  4800 &   3.1    &  1.50  &  ---  &   ---    &  ---    \\       
Her 35  &  5150 &   3.3    &  1.55  &  ---  &   ---    &  ---    \\        
Her 82  &  5150 &   3.3    &  1.30  &  ---  &   ---    &  ---    \\       
Her 87  &  5100 &   3.2    &  1.30  &  5000 &   2.9    &  1.80   \\       
Her 90  &  4700 &   3.1    &  1.60  &  ---  &   ---    &  ---    \\       
Her 144 &  5100 &   3.2    &  1.65  &  5200 &   3.2    &  2.00   \\       
Her 170 &  5150 &   3.0    &  1.20  &  ---  &   ---    &  ---    \\        
Her 176 &  5100 &   3.0    &  1.00  &  5200 &   3.0    &  2.00   \\       
Her 228 &  5150 &   3.2    &  1.40  &  5000 &   2.9    &  1.80   \\       
Her 249 &  5150 &   3.3    &  1.30  &  5000 &   2.9    &  1.80   \\       
Her 296 &  5200 &   3.3    &  1.20  &  5000 &   2.9    &  1.80   \\       
Her 397 &  5200 &   3.4    &  1.40  &  5000 &   3.0    &  1.80   \\       
\hline
\end{tabular}
\end{center}
\end{table}

\subsection[]{Line lists and atomic data}\label{subsec:atomic_data}

\begin{figure*}
\begin{center}
\begin{minipage}{170mm}
\hspace{-0.9cm}
$\begin{array}{ccc}
\includegraphics[width=2.5in]{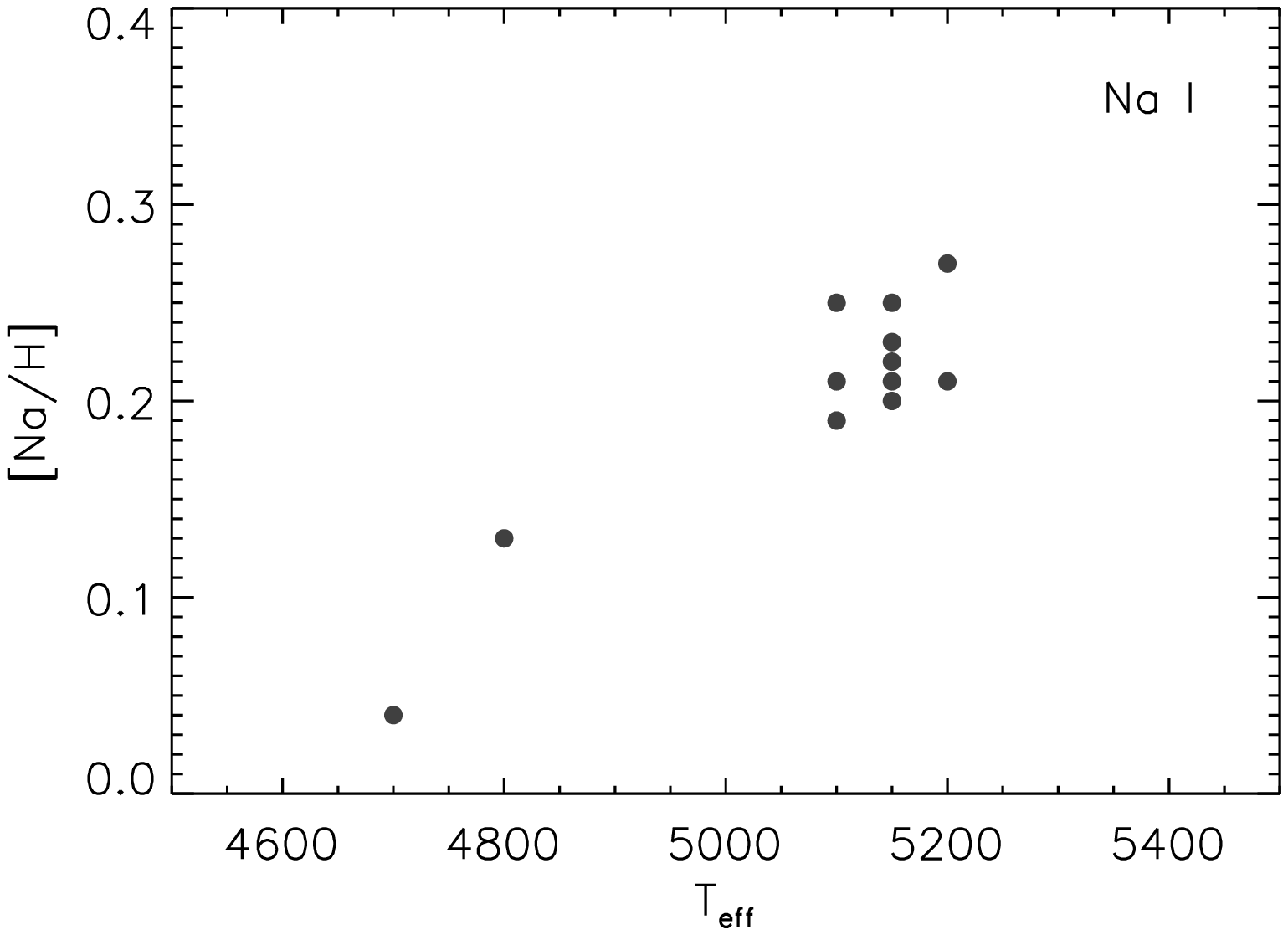} \hspace{-0.7cm} & 
\includegraphics[width=2.5in]{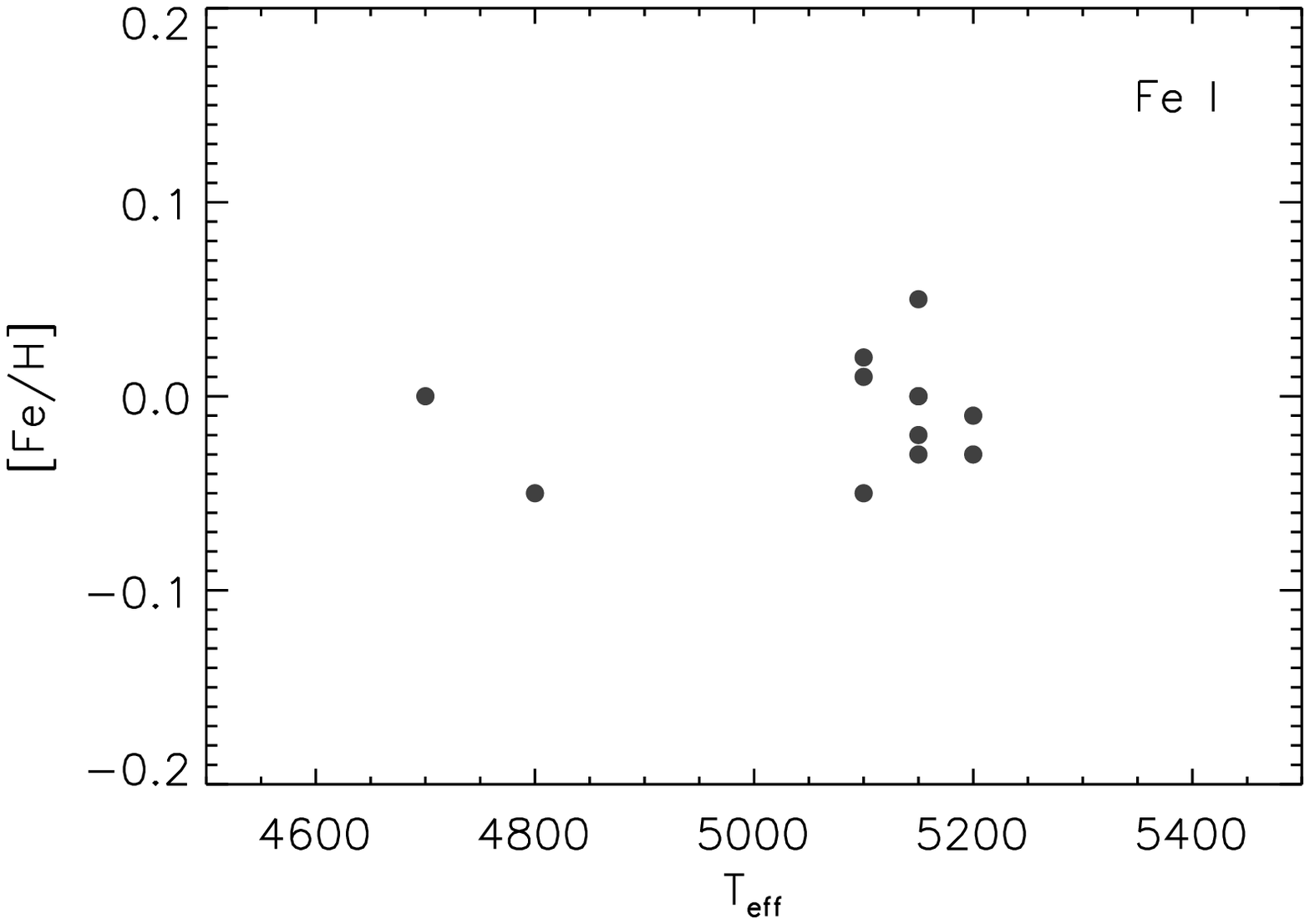} \hspace{-0.7cm} & 
\includegraphics[width=2.5in]{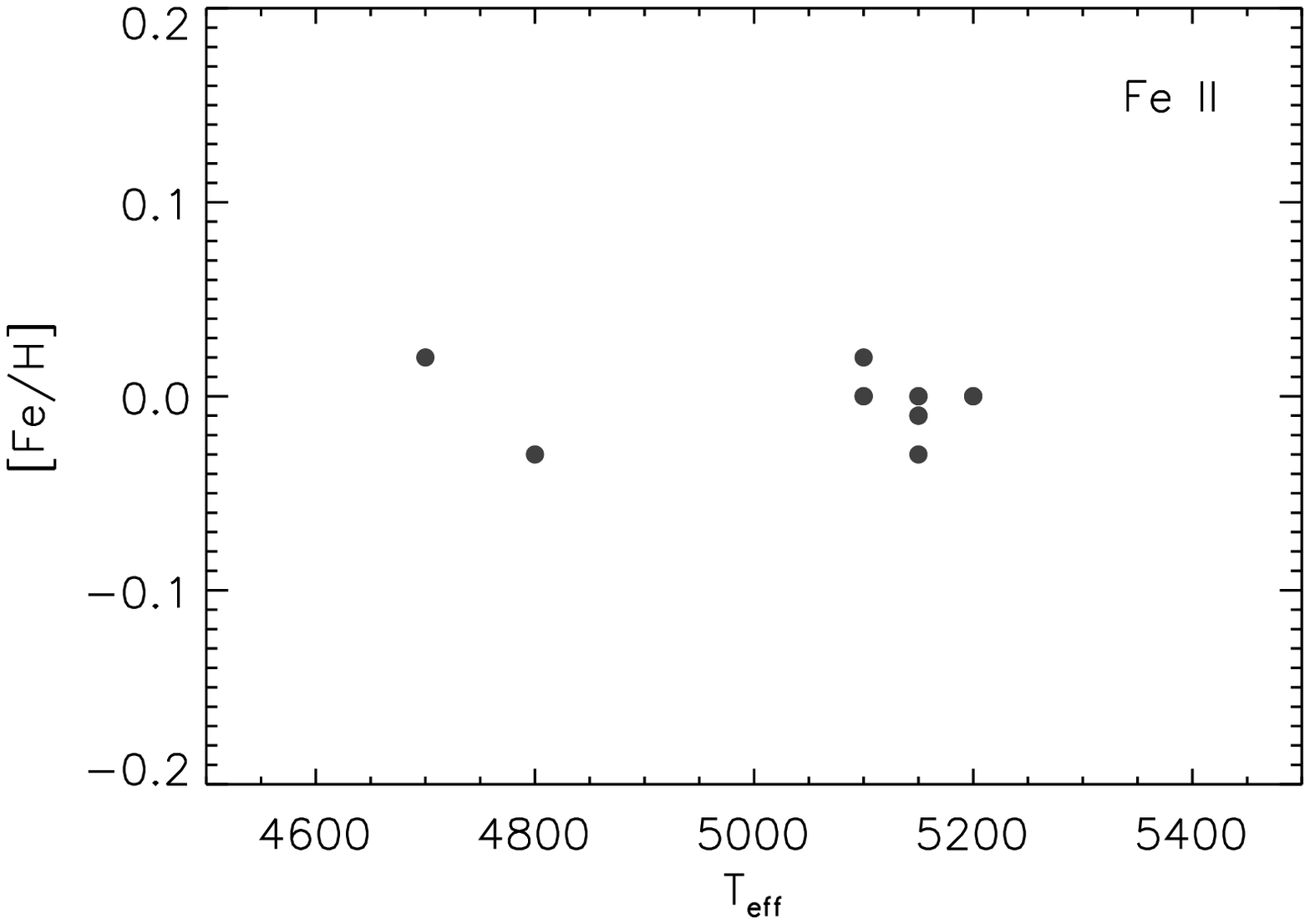} 
\end{array}$
\caption{[X/H] ratios as a function of spectroscopically determined effective temperature: [Na/H] (left-hand panel), [Fe\,{\sc i}/H] (middle panel), [Fe\,{\sc ii}/H] (right-hand panel).}\label{fig:teff}
\end{minipage}
\end{center}
\end{figure*}

\begin{table*}
\begin{center}
\caption{Fe, Na, and Al abundances in [X/H] format. $N$ is the total number of lines used and $\sigma$ is the total measurement uncertainty including the line-to-line scatter.\label{table:EW_analysis_1}}
\begin{tabular}{lccccccccccccccc}
\hline
& \multicolumn{3}{c}{Fe\,{\sc i}} &$\,$& \multicolumn{3}{c}{Fe\,{\sc ii}} &$\,$& \multicolumn{3}{c}{Na\,{\sc i}} &$\,$& \multicolumn{3}{c}{Al\,{\sc i}} \\[0.05cm] 
\\[-0.2cm]
Star                & [X/H] & $N$&$\sigma$&$\,$& [X/H] & $N$&$\sigma$&$\,$& [X/H] & $N$&$\sigma$&$\,$& [X/H] & $N$&$\sigma$\\
\hline
Her 6               & -0.05 & 24 &  0.09  &$\,$& -0.03 &  6 &  0.13  &$\,$&  0.13 &  4 &  0.10  &$\,$&  0.10 &  2 &  0.07  \\
Her 35              & -0.02 & 34 &  0.10  &$\,$& -0.03 &  4 &  0.10  &$\,$&  0.20 &  4 &  0.08  &$\,$&  0.11 &  2 &  0.05  \\
Her 82              & -0.03 & 35 &  0.11  &$\,$&  0.00 &  6 &  0.09  &$\,$&  0.21 &  4 &  0.08  &$\,$&  0.12 &  2 &  0.07  \\
Her 87              &  0.01 & 38 &  0.12  &$\,$&  0.00 &  5 &  0.12  &$\,$&  0.19 &  4 &  0.08  &$\,$&  0.13 &  2 &  0.06  \\
Her 90              &  0.00 & 26 &  0.07  &$\,$&  0.02 &  5 &  0.14  &$\,$&  0.04 &  4 &  0.09  &$\,$&  0.07 &  1 &  0.08  \\
Her 144             & -0.05 & 36 &  0.09  &$\,$&  0.00 &  7 &  0.10  &$\,$&  0.25 &  4 &  0.08  &$\,$&  0.10 &  2 &  0.07  \\
Her 170             &  0.00 & 30 &  0.11  &$\,$& -0.01 & 10 &  0.11  &$\,$&  0.22 &  4 &  0.08  &$\,$&  0.09 &  2 &  0.06  \\
Her 176             &  0.02 & 31 &  0.10  &$\,$&  0.02 &  4 &  0.10  &$\,$&  0.21 &  4 &  0.09  &$\,$&  0.12 &  2 &  0.07  \\
Her 228             &  0.00 & 31 &  0.10  &$\,$&  0.00 &  6 &  0.10  &$\,$&  0.23 &  4 &  0.09  &$\,$&  0.06 &  2 &  0.07  \\
Her 249             &  0.05 & 32 &  0.09  &$\,$& -0.01 &  8 &  0.10  &$\,$&  0.25 &  4 &  0.08  &$\,$&  0.12 &  2 &  0.05  \\
Her 296             & -0.01 & 32 &  0.09  &$\,$&  0.00 &  4 &  0.11  &$\,$&  0.21 &  4 &  0.09  &$\,$&  0.11 &  2 &  0.06  \\
Her 397             & -0.03 & 29 &  0.08  &$\,$&  0.00 &  3 &  0.10  &$\,$&  0.27 &  4 &  0.09  &$\,$&  0.16 &  2 &  0.05  \\
\\
Mean                & -0.01 & ---&  0.10  &$\,$&  0.00 & ---&  0.11  &$\,$&  0.21 & ---&  0.10  &$\,$&  0.11 & ---&  0.07  \\
\hline
\end{tabular}
\end{center}
\end{table*}

To carry out the spectroscopic analysis, we employ atomic lines for Na, Al, Mg, Si, Ca, Ti, Cr, Ni, Fe and Zn, which are a subset of the line lists of \citet*{ben03}, to enable abundance comparison. Lines were selected to  minimize line-blending, and during the analysis all blended lines were discarded from further use. The full line lists, atomic data adopted and equivalent widths (EWs) measured can be found in Appendix~\ref{appendix:equivalent_widths}. To facilitate the comparison between our study and previous studies, we use the solar abundances from \citet{gre98}. Different choice of solar abundances shall not have significant impact on our conclusion because the solar abundances for elements $Z \geq 12$ are largely the same.

\subsection[]{Atmospheric parameters}\label{subsec:atmospheric_parameter}

We determine the atmospheric parameters, e.g. effective temperature (\teff), surface gravity ($\log g$) and microturbulence ($v_t$) using {\sc moog} (Sneden, 2010 version) program and Kurucz models assuming local thermodynamic equilibrium (LTE). The atmospheric parameters grid was interpolated using \citet{cas94}. We determine the spectroscopic atmospheric parameters by an iterative manner assuming the photometric effective temperature to be initial effective temperature. The photometric effective temperature is calculated based on \citet*{alo99} calibration. We minimize the differences between [{Fe\,{\sc i}}/H] and [{Fe\,{\sc ii}}/H] (i.e. ionization equilibrium) to derive $\log g$, eliminate the slope of abundances (in log eps) as a function of excitation potential (EP) to derive \teff\, and reduced EW in log scale to determine $v_t$. 

We found that photometric effective temperatures are systematically lower than the spectroscopic effective temperature, of the order of $300$ K, consistent with the results from \citet{jac07}. This is most likely due to not taking into account extinction in photometry. It has also been reported that IC 4756 is subject to variable extinction \citep{sch78,smi83}.

\subsection[]{Equivalent width}\label{subsec:equivalent-width}
\label{subsection:ew}

We measure the atomic line EWs by fitting a Gaussian profile interactively with {\sc iraf splot} package. The elemental abundances were then calculated using {\sc moog} adopting the Kurucz models with the best-fitting atmospheric parameters as determined spectroscopically. We make sure that, for each element, all lines give the same abundances within 0.1 dex, and discard blended lines carefully. The final lines and EWs that we use to deduce the abundances can be found in Appendix~\ref{appendix:equivalent_widths}.

\begin{table*}
\begin{center}
\caption{Same as Table \ref{table:EW_analysis_1}, but for Mg, Si, Ca and Ti abundances: $\alpha$-elements. \label{table:EW_analysis_2}}
\begin{tabular}{lccccccccccccccccccc}
\hline 
& \multicolumn{3}{c}{Mg\,{\sc i}} &$\,$& \multicolumn{3}{c}{Si\,{\sc i}} &$\,$& \multicolumn{3}{c}{Ca\,{\sc i}} &$\,$& \multicolumn{3}{c}{Ti\,{\sc i}} &$\,$& \multicolumn{3}{c}{Ti\,{\sc ii}} \\[0.05cm]
\\[-0.2cm]
Star                & [X/H] & $N$&$\sigma$&$\,$& [X/H] & $N$&$\sigma$&$\,$& [X/H] & $N$&$\sigma$&$\,$& [X/H] & $N$&$\sigma$&$\,$& [X/H] & $N$&$\sigma$\\
\hline
Her 6               &  0.11 &  1 &  0.09  &$\,$&  0.15 & 11 &  0.08  &$\,$& -0.01 & 15 &  0.13  &$\,$&  0.02 & 13 &  0.18  &$\,$&  0.27 &  7 &  0.10  \\
Her 35              &  0.05 &  1 &  0.06  &$\,$&  0.14 & 13 &  0.06  &$\,$&  0.06 & 17 &  0.12  &$\,$& -0.05 & 17 &  0.16  &$\,$&  0.27 &  7 &  0.07  \\
Her 82              &  0.16 &  1 &  0.08  &$\,$&  0.14 & 13 &  0.06  &$\,$&  0.03 & 17 &  0.10  &$\,$&  0.00 & 18 &  0.15  &$\,$&  0.30 &  8 &  0.06  \\
Her 87              &  0.15 &  1 &  0.09  &$\,$&  0.16 & 12 &  0.05  &$\,$&  0.03 & 16 &  0.13  &$\,$& -0.01 & 14 &  0.17  &$\,$&  0.28 &  8 &  0.10  \\
Her 90              &  0.08 &  1 &  0.05  &$\,$&  0.12 & 11 &  0.09  &$\,$&  0.01 & 12 &  0.13  &$\,$& -0.02 & 15 &  0.17  &$\,$&  0.29 &  4 &  0.06  \\
Her 144             &  0.09 &  1 &  0.08  &$\,$&  0.04 & 13 &  0.06  &$\,$&  0.01 & 16 &  0.12  &$\,$& -0.04 & 16 &  0.14  &$\,$&  0.27 &  8 &  0.07  \\
Her 170             &  0.15 &  1 &  0.08  &$\,$&  0.02 & 13 &  0.06  &$\,$&  0.03 & 16 &  0.13  &$\,$& -0.03 & 17 &  0.16  &$\,$&  0.28 &  8 &  0.10  \\
Her 176             &  0.13 &  1 &  0.07  &$\,$&  0.07 & 13 &  0.06  &$\,$&  0.03 & 17 &  0.11  &$\,$&  0.01 & 17 &  0.16  &$\,$&  0.28 &  6 &  0.09  \\
Her 228             &  0.15 &  1 &  0.07  &$\,$&  0.14 & 13 &  0.04  &$\,$&  0.08 & 17 &  0.11  &$\,$& -0.01 & 18 &  0.14  &$\,$&  0.29 &  7 &  0.07  \\
Her 249             &  0.13 &  1 &  0.08  &$\,$&  0.13 & 13 &  0.08  &$\,$&  0.05 & 17 &  0.11  &$\,$&  0.03 & 17 &  0.16  &$\,$&  0.29 &  7 &  0.10  \\
Her 296             &  0.05 &  1 &  0.08  &$\,$&  0.10 & 13 &  0.05  &$\,$&  0.03 & 17 &  0.12  &$\,$& -0.01 & 18 &  0.15  &$\,$&  0.30 &  8 &  0.10  \\
Her 397             &  0.14 &  1 &  0.07  &$\,$&  0.17 & 13 &  0.06  &$\,$&  0.09 & 17 &  0.11  &$\,$&  0.00 & 18 &  0.16  &$\,$&  0.29 &  8 &  0.08  \\
\\
Mean                &  0.11 & ---&  0.08  &$\,$&  0.12 & ---&  0.08  &$\,$&  0.04 & ---&  0.12  &$\,$& -0.01 & ---&  0.16  &$\,$&  0.28 & ---&  0.08  \\
\hline
\end{tabular}
\end{center}
\end{table*}

\begin{table*}
\begin{center}
\caption{Same as Table \ref{table:EW_analysis_1}, but for Cr, Ni, Zn and Ba abundances: Fe-peak elements and neutron-capture elements.\label{table:EW_analysis_3}}
\begin{tabular}{lccccccccccccccccccc}
\hline 
& \multicolumn{3}{c}{Cr\,{\sc i}} &$\,$& \multicolumn{3}{c}{Cr\,{\sc ii}} &$\,$& \multicolumn{3}{c}{Ni\,{\sc i}} &$\,$& \multicolumn{3}{c}{Zn\,{\sc i}} &$\,$& \multicolumn{3}{c}{Ba\,{\sc i}}  \\[0.05cm]
\\[-0.2cm]
Star                & [X/H] & $N$&$\sigma$&$\,$& [X/H] & $N$&$\sigma$&$\,$& [X/H] & $N$&$\sigma$&$\,$& [X/H] & $N$&$\sigma$&$\,$& [X/H] & $N$&$\sigma$\\
\hline
Her 6               & -0.02 &  5 &  0.13  &$\,$&  0.26 &  4 &  0.11  &$\,$&  0.03 & 42 &  0.09 &$\,$&  0.03 &  2 &  0.09  &$\,$& -0.03 &  1 &  0.10  \\
Her 35              &  0.01 &  5 &  0.11  &$\,$&  0.28 &  5 &  0.09  &$\,$&  0.03 & 43 &  0.06  &$\,$&  0.04 &  2 &  0.07  &$\,$&  0.00 &  1 &  0.10  \\
Her 82              & -0.03 &  5 &  0.10  &$\,$&  0.25 &  5 &  0.07  &$\,$&  0.00 & 43 &  0.07  &$\,$&  0.05 &  2 &  0.05  &$\,$& -0.05 &  1 &  0.10  \\
Her 87              &  0.01 &  5 &  0.11  &$\,$&  0.27 &  5 &  0.09  &$\,$&  0.02 & 42 &  0.08  &$\,$&  0.04 &  2 &  0.09  &$\,$&  0.03 &  1 &  0.10  \\
Her 90              &  0.07 &  4 &  0.13  &$\,$&  0.26 &  4 &  0.11  &$\,$&  0.05 & 27 &  0.07  &$\,$&  0.09 &  1 &  0.11  &$\,$&  0.00 &  1 &  0.10  \\
Her 144             & -0.01 &  4 &  0.10  &$\,$&  0.20 &  5 &  0.07  &$\,$& -0.03 & 43 &  0.07  &$\,$&  0.04 &  2 &  0.07  &$\,$& -0.05 &  1 &  0.10  \\
Her 170             & -0.01 &  5 &  0.11  &$\,$&  0.16 &  5 &  0.10  &$\,$&  0.00 & 42 &  0.10  &$\,$&  0.01 &  2 &  0.08  &$\,$& -0.01 &  1 &  0.10  \\
Her 176             & -0.01 &  5 &  0.10  &$\,$&  0.23 &  5 &  0.09  &$\,$&  0.02 & 42 &  0.09  &$\,$&  0.04 &  2 &  0.07  &$\,$&  0.00 &  1 &  0.10  \\
Her 228             &  0.02 &  4 &  0.11  &$\,$&  0.21 &  5 &  0.08  &$\,$&  0.03 & 41 &  0.09  &$\,$&  0.07 &  2 &  0.06  &$\,$& -0.02 &  1 &  0.10  \\
Her 249             &  0.06 &  5 &  0.11  &$\,$&  0.27 &  5 &  0.08  &$\,$&  0.04 & 43 &  0.09  &$\,$&  0.07 &  2 &  0.07  &$\,$&  0.00 &  1 &  0.10  \\
Her 296             &  0.02 &  5 &  0.12  &$\,$&  0.24 &  5 &  0.08  &$\,$&  0.05 & 42 &  0.09  &$\,$&  0.09 &  2 &  0.07  &$\,$&  0.05 &  1 &  0.10  \\
Her 397             &  0.07 &  5 &  0.10  &$\,$&  0.26 &  5 &  0.08  &$\,$&  0.00 & 43 &  0.08  &$\,$&  0.05 &  2 &  0.06  &$\,$&  0.02 &  1 &  0.10  \\
\\
Mean                &  0.01 & ---&  0.12  &$\,$&  0.24 & ---&  0.09  &$\,$&  0.02 & ---&  0.08  &$\,$&  0.05 & ---&  0.08  &$\,$& -0.01 & ---&  0.10  \\
\hline
\end{tabular}
\end{center}
\end{table*}

We do not find any correlation between [Fe/H] or [X/Fe] versus the effective temperature as shown in Fig.~\ref{fig:teff}, except for Na. The Spearman's $\rho$ test or Kendall's $\tau$ test always give $p$-value larger than 0.1, where $p$-value is the probability of false positive correlation. The lack of correlation of both [Fe\,{\sc i}/H] and [Fe\,{\sc ii}/H] with \teff~ suggests that our estimated atmospheric parameters are consistent in itself. However, for Na, both [Na/H] and [Na/Fe] show strong correlation with the effective temperature, with $p$-value less than 0.03, i.e. less than $3$ per cent of chance that we will get false positive correlation. This trend might be due to unquantified analysis effects such as unknown line-blends or non-LTE (NLTE) effects \citep{adi12}. Our Na analysis is based on measurement of the 5682/5688 doublet and the 6154/6160 doublet. We find that both doublets give the similar result with no difference seen if we considered only one set of the lines. It is still a possibility that NLTE effects can explain the trend, where a correction would decrease the enhanced Na abundances \citep*{shi, lind}. This Na trend could also be due to internal mixing in the stars and we will discuss in details in Section~\ref{sec:results_and_discussion}.

We estimate the total uncertainty in elemental abundances for each star individually by combining in quadrature the line-to-line scatter for each element and the uncertainties due to the determination of three atmospheric parameters: effective temperature, surface gravity and microturbulence. For microturbulence, we vary $v_t$ until the abundances versus the EW in log scale has a slope greater than 0.05, and take this range as $1\sigma$. We assume the $1\sigma$ uncertainties in effective temperature to be 100 K. We choose this range because our results for various independent trials of determining effective temperature are always within this range. We vary the surface gravity until the differences between [{Fe\,{\sc i}}/H] and [{Fe\,{\sc ii}}/H] are greater than 0.05 dex. Since there are non-trivial correlations between these three uncertainties, combining in quadrature will give us a conservative upper-limit of the total uncertainty.

\subsection[]{Spectral synthesis}\label{subsec:spectrum_synthesis}

\begin{figure*}
\begin{center}
\begin{minipage}{170mm}
\hspace{-0.9cm}
$\begin{array}{ccc}
\includegraphics[width=2.5in]{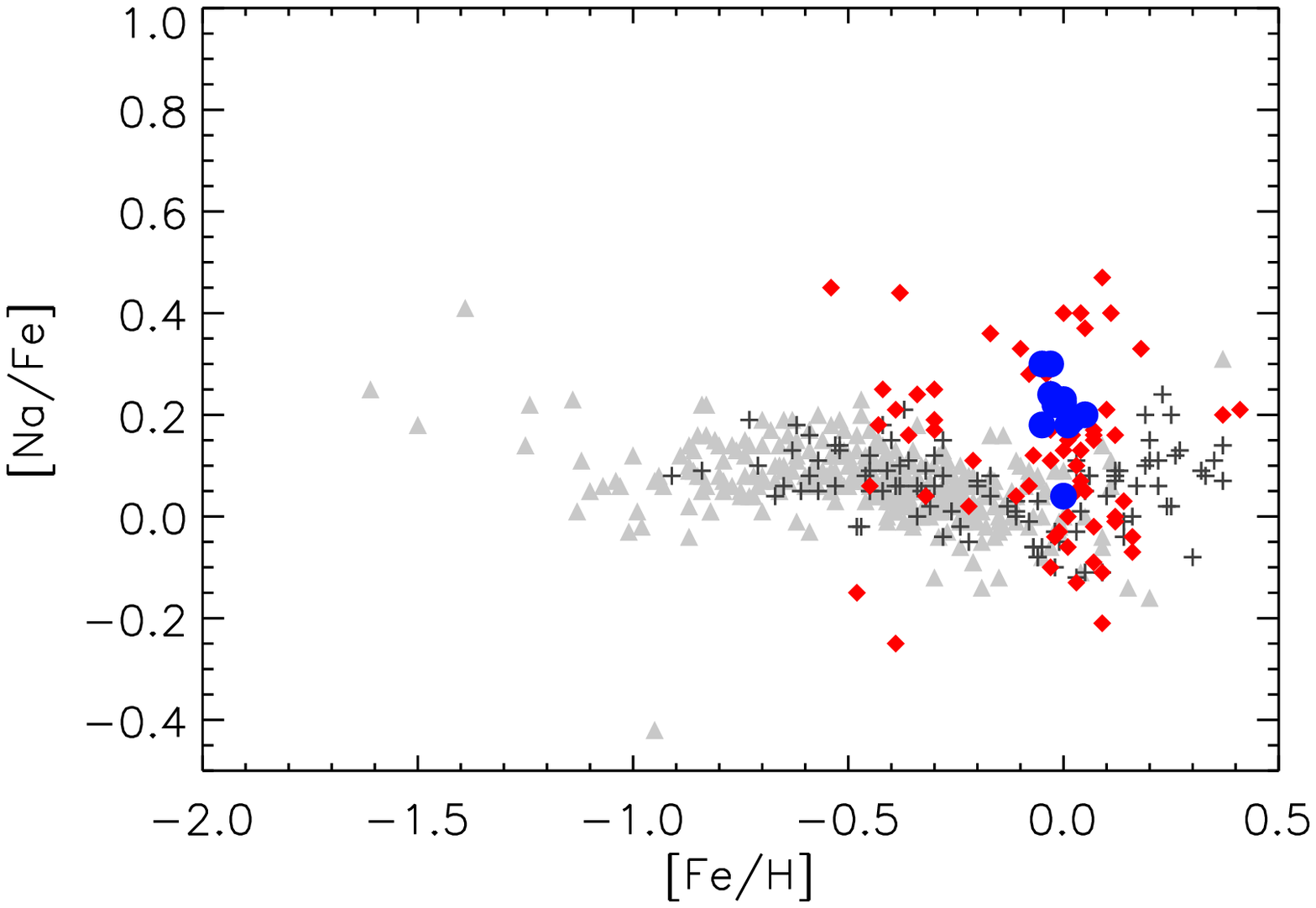} \hspace{-0.7cm} & 
\includegraphics[width=2.5in]{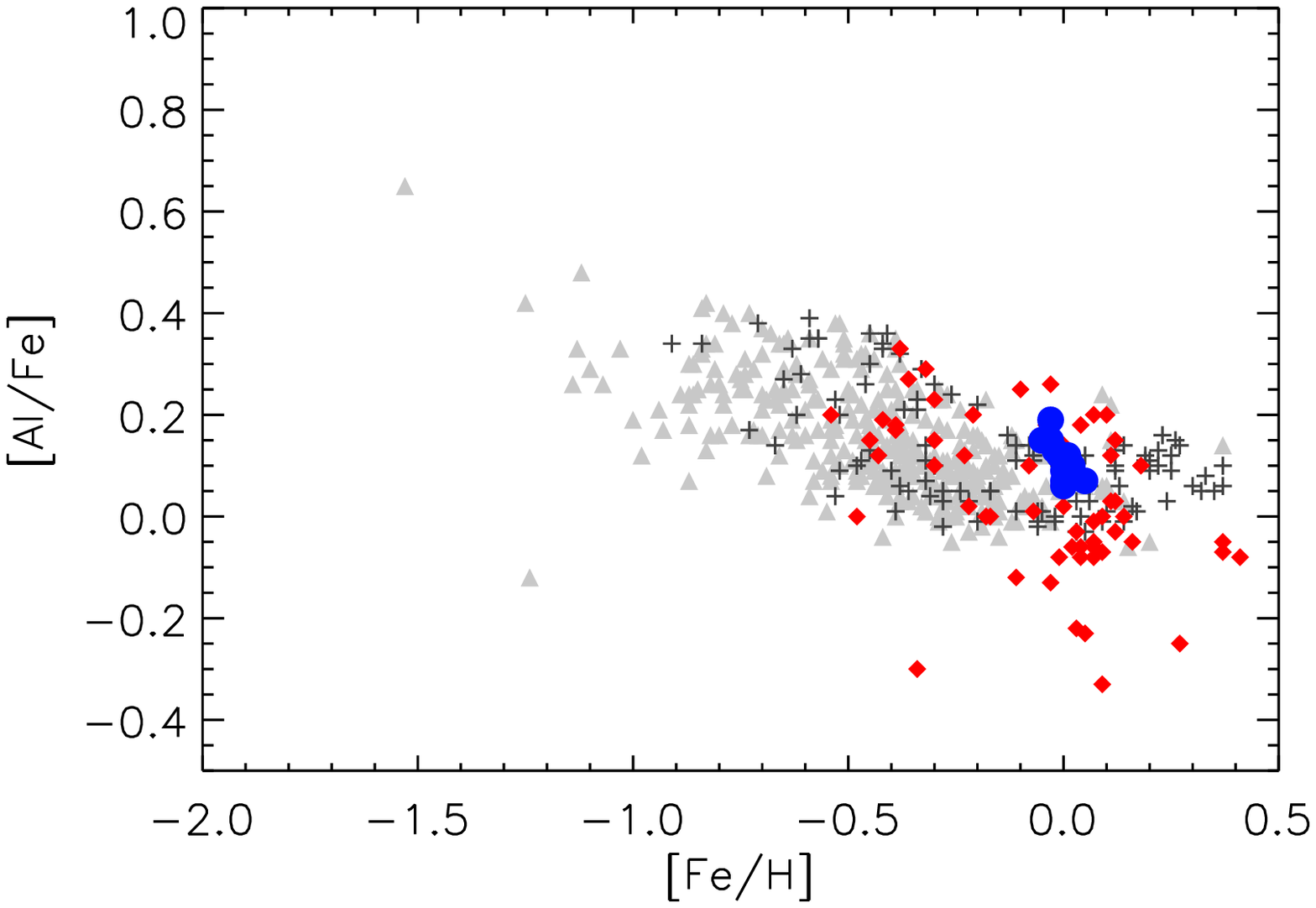} \hspace{-0.7cm} & 
\includegraphics[width=2.5in]{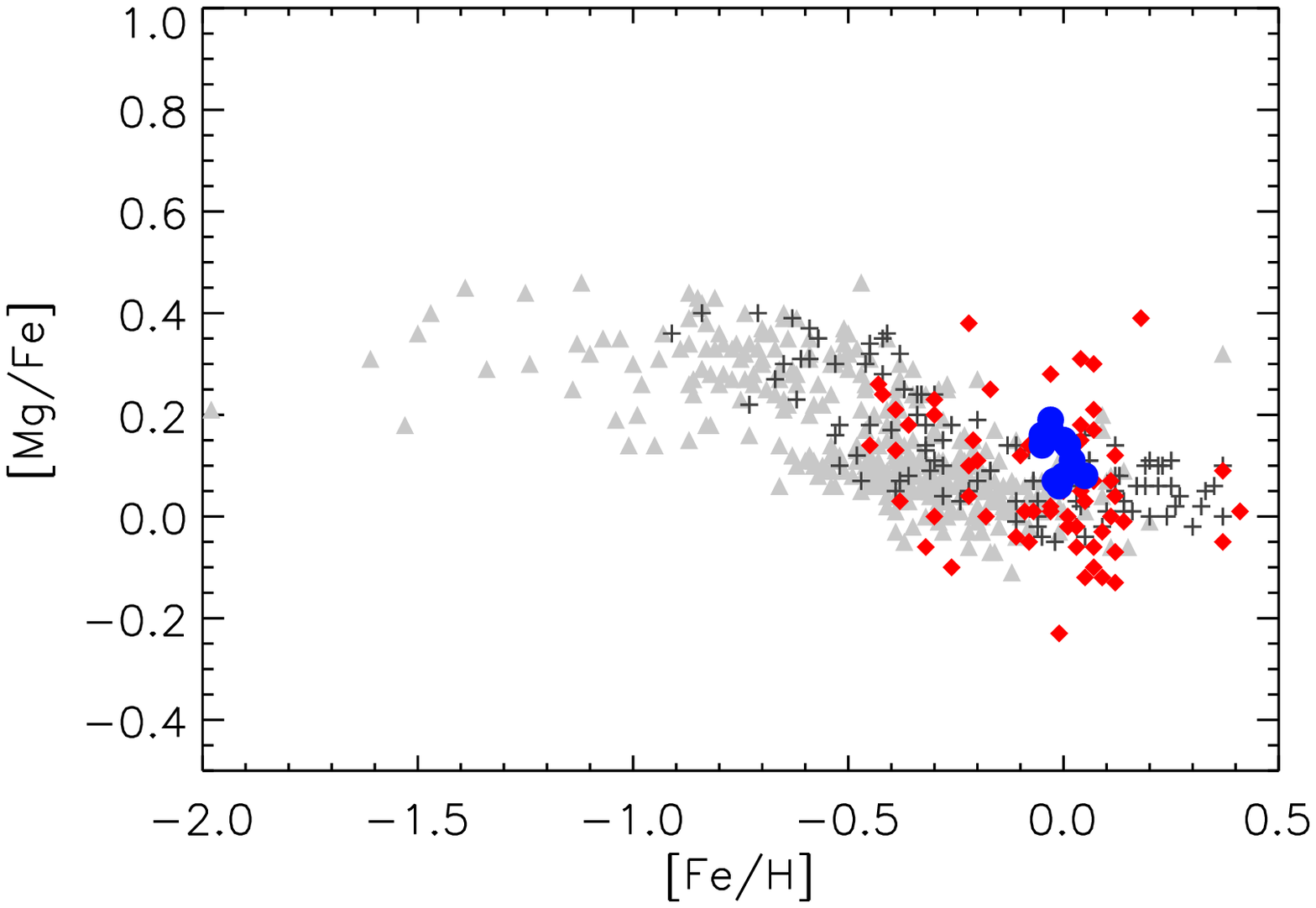} \vspace{-0.4cm}\\ 
\includegraphics[width=2.5in]{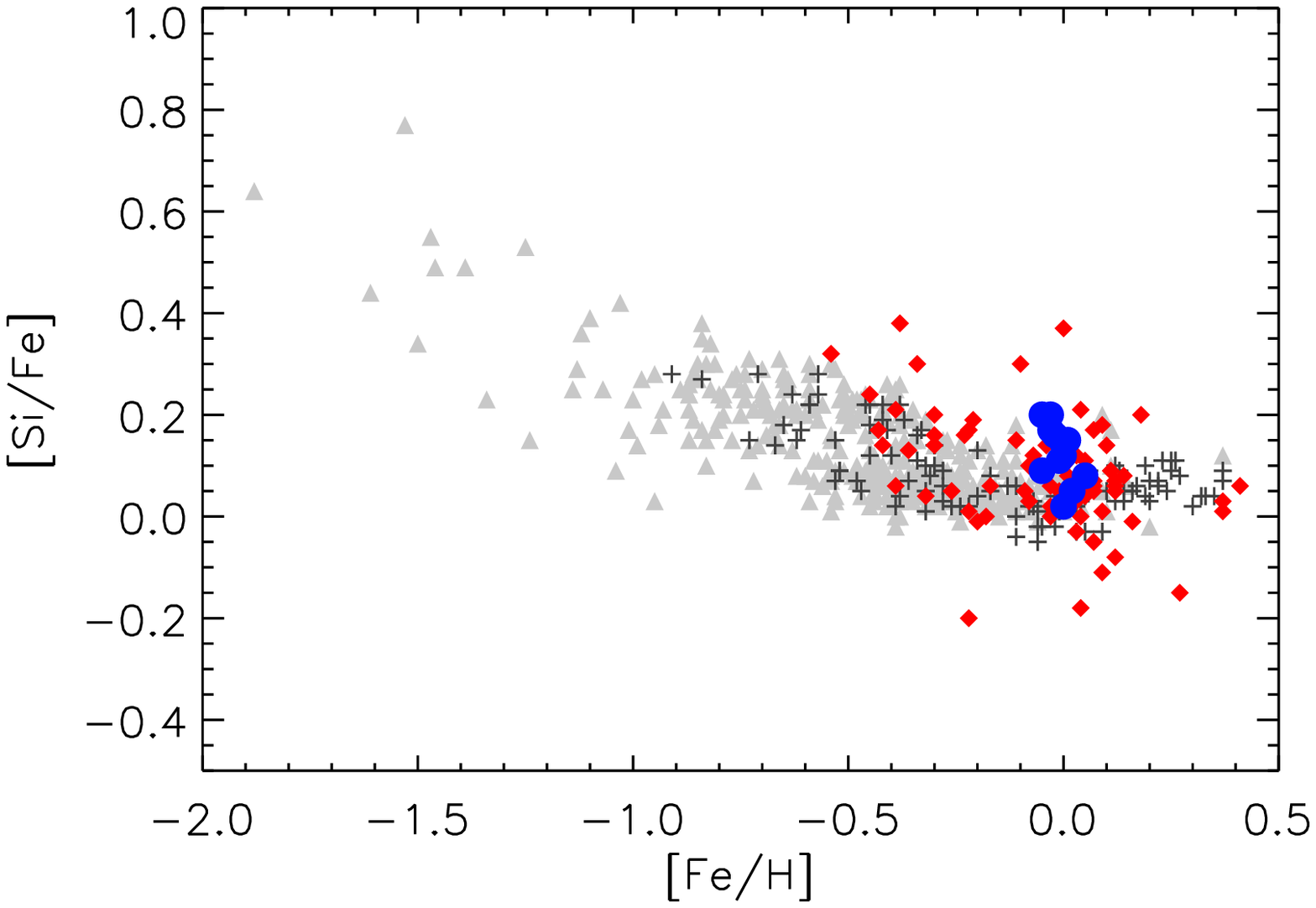} \hspace{-0.7cm} & 
\includegraphics[width=2.5in]{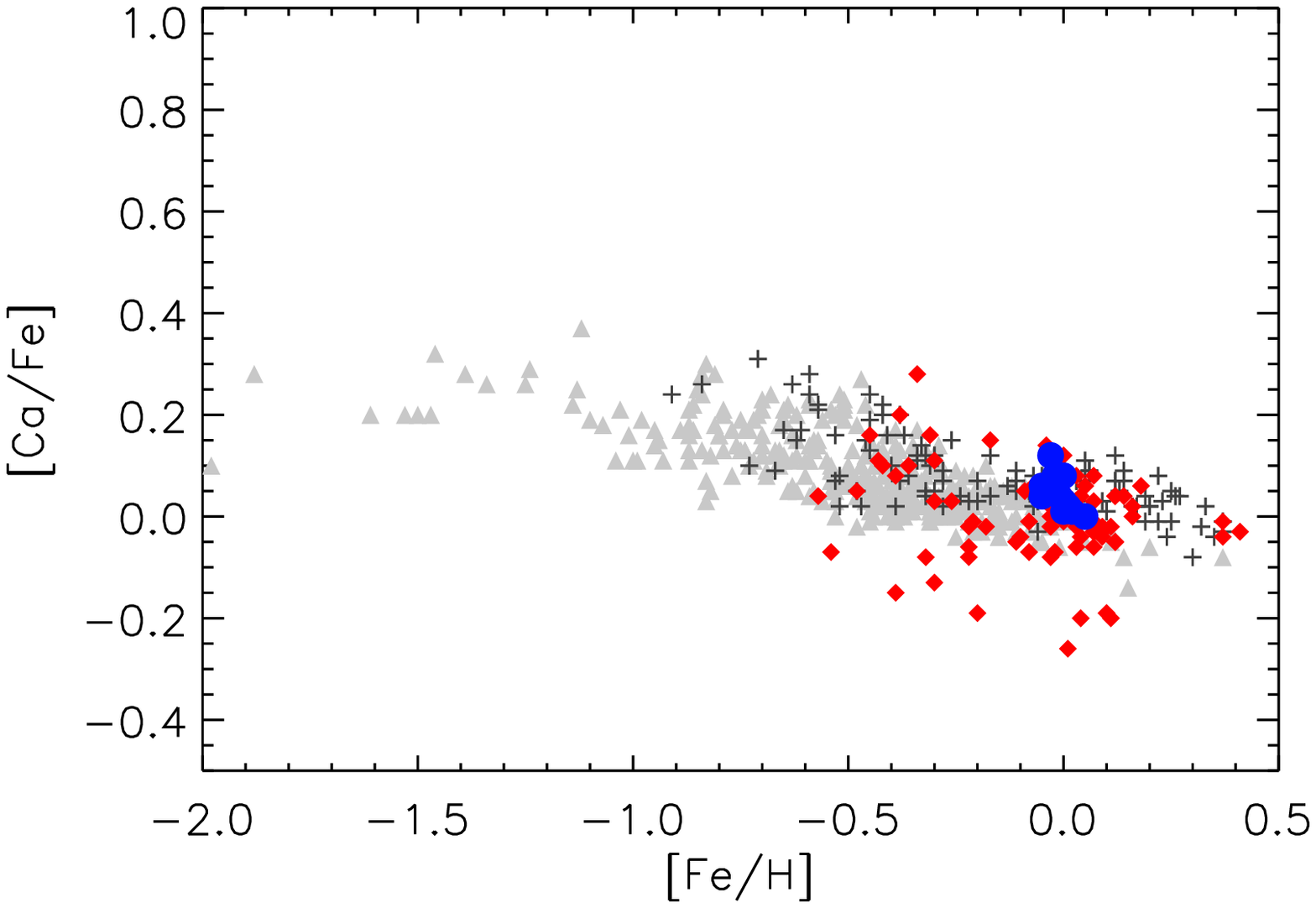} \hspace{-0.7cm} & 
\includegraphics[width=2.5in]{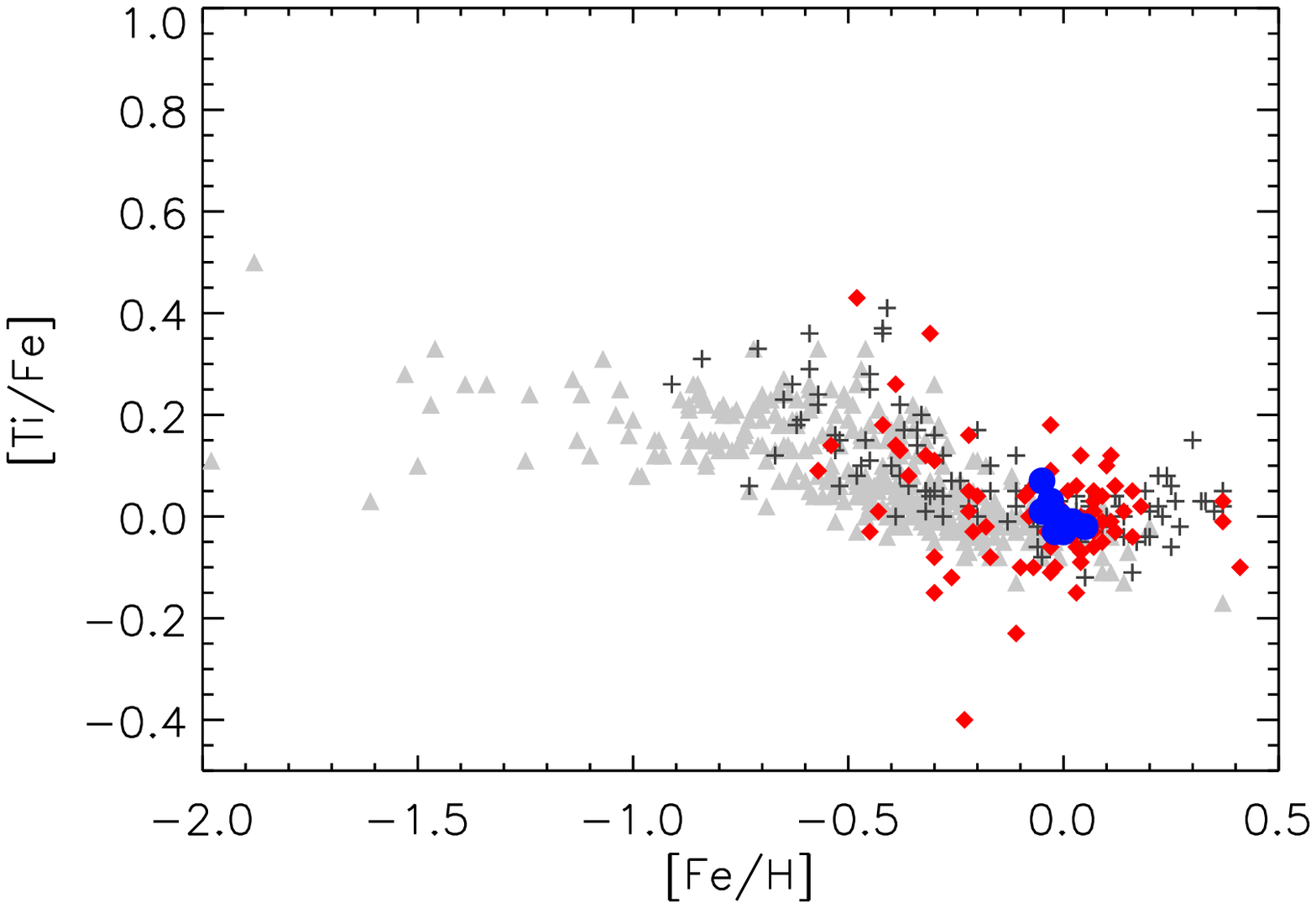} \vspace{-0.4cm}\\
\includegraphics[width=2.5in]{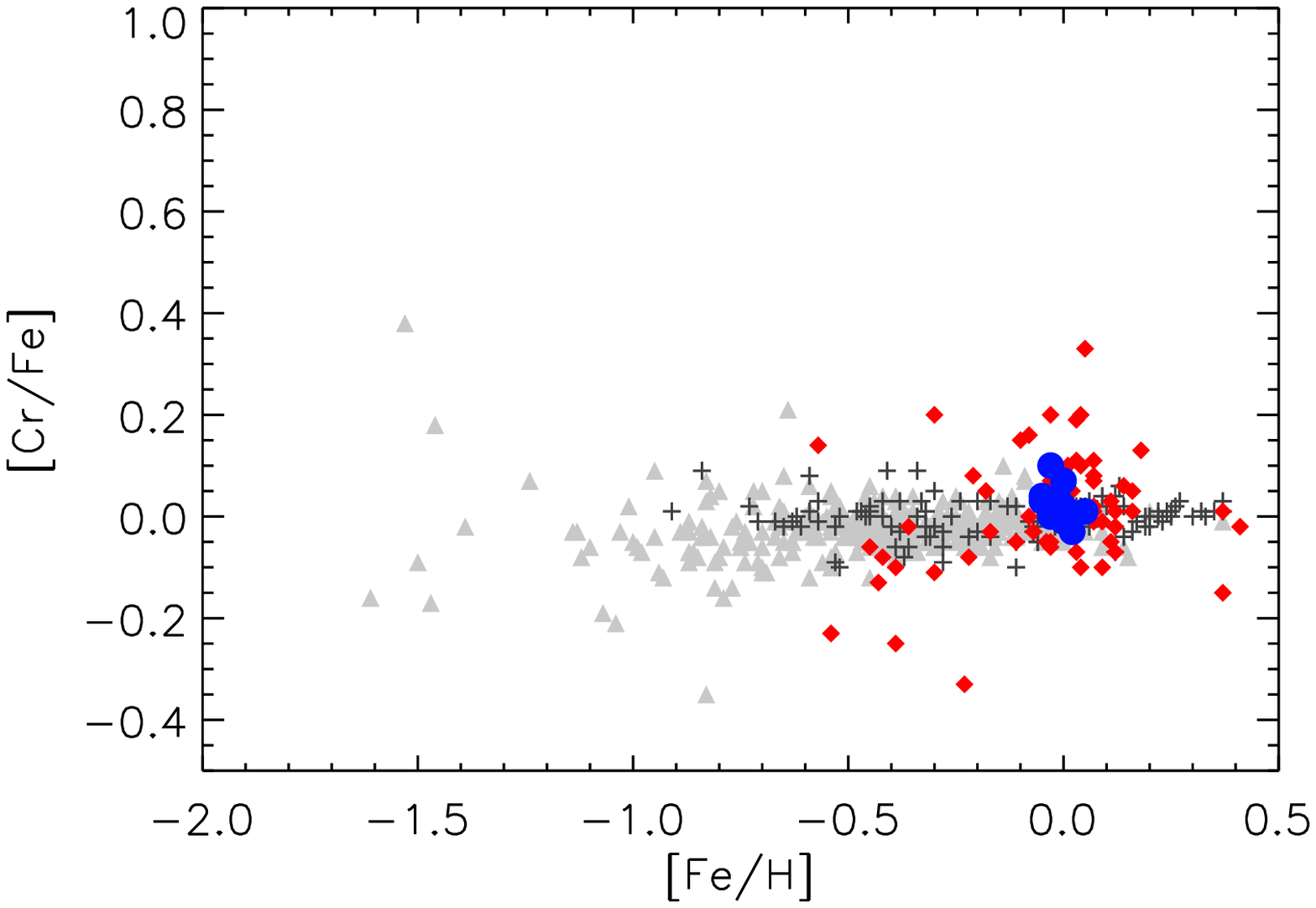} \hspace{-0.7cm} & 
\includegraphics[width=2.5in]{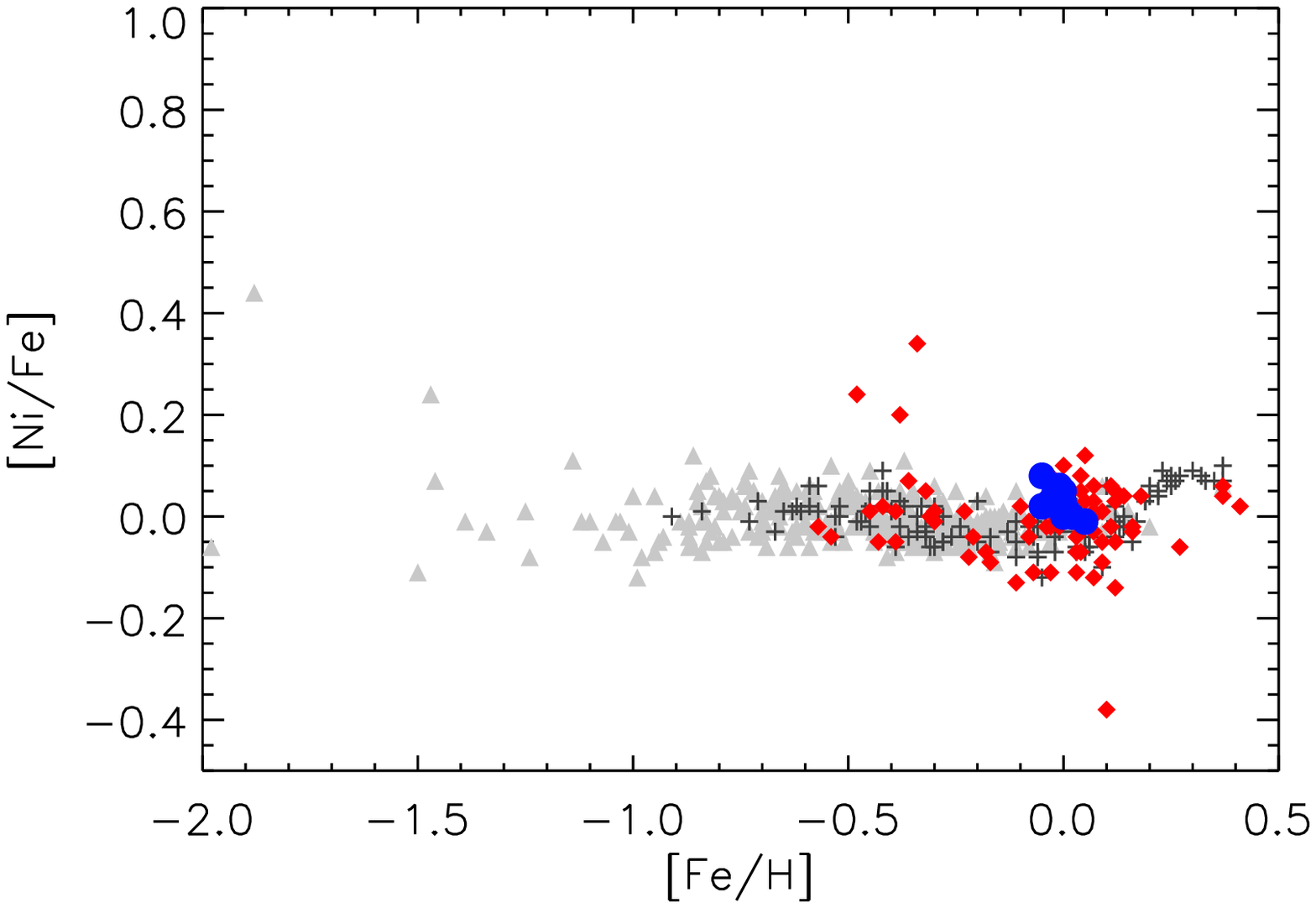} \hspace{-0.7cm} &
\includegraphics[width=2.5in]{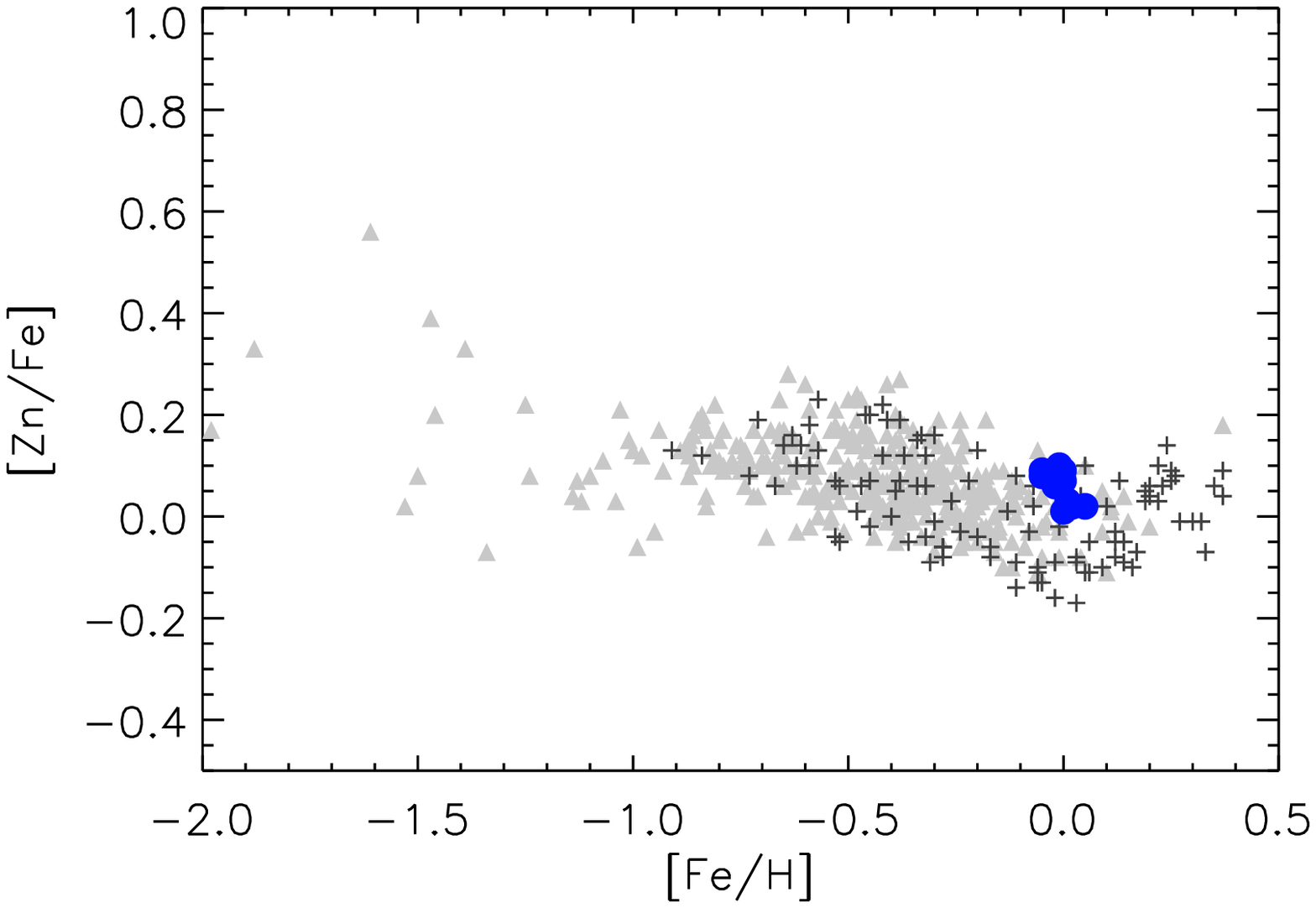} \vspace{-0.4cm}\\
&
\includegraphics[width=2.5in]{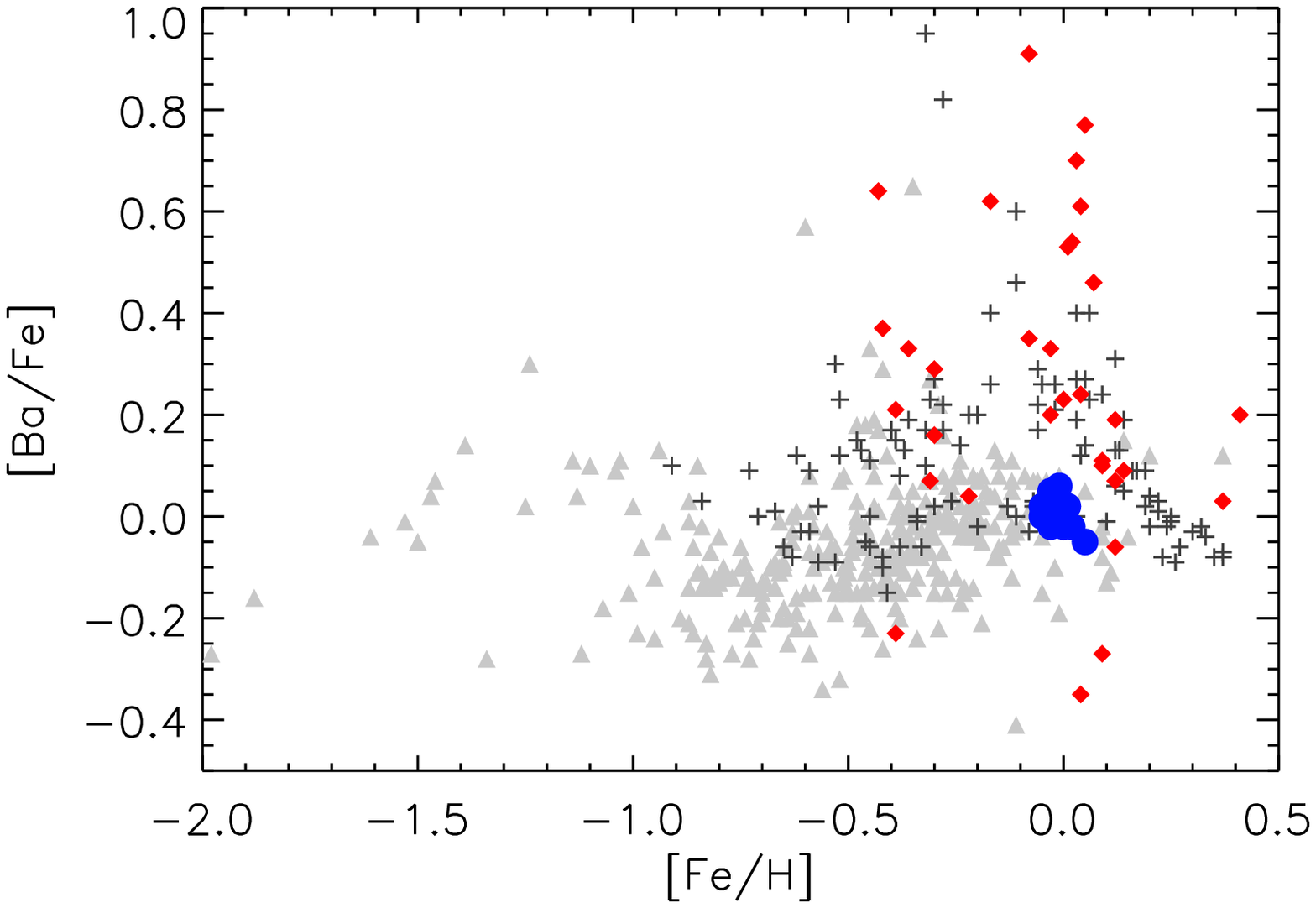} \hspace{-0.7cm} &
\end{array}$
\caption{[X/Fe] as a function of [Fe/H] for IC 4756 subgiants of this study (filled blue circles), other open clusters from \citet{car11} (red diamonds) and solar neighbourhood disc stars from \citet{red03}, \citet*{red06} (grey triangles) and \citet{ben03} (black plus). All samples assume 1D LTE stellar model.}\label{fig:xfe_feh}
\end{minipage}
\end{center}
\end{figure*}

For Ba, we target the Ba\,{\sc ii} line at 5853 \AA\ taking into account hyperfine structure and isotopic splitting adopting a solar isotopic ratio. This line is strong, unblended and not thought to be subject to NLTE effects \citep[e.g.][]{mas07}. Spectral syntheses of the respective regions were carried out using the {\sc moog synth} driver. Assuming the uncertainty on microturbulence to be 0.10 (which is the upper limit for our analysis as described in Section~\ref{subsection:ew}), typical total uncertainty of the elemental abundance is of the order of 0.1 dex. The uncertainty is dominated by the uncertainty in microturbulence (about 0.08 dex), while the uncertainty due to \teff\ and $\log g$ is about 0.02 dex and uncertainty in the synthesis fit is about 0.05 dex. 

\subsection[]{Cluster mean abundances}\label{subsec:mean_abundances}

In Table~\ref{table:EW_analysis_1}--\ref{table:EW_analysis_3}, we list our abundance results in [X/H] for all studied elements. In each table, $N$ is the total number of lines used and the $\sigma$ gives the total uncertainty per star per element. Except Na, all elemental abundances show little star-to-star dispersion. However, we wish to be conservative in calculating the cluster mean abundances {\it by taking into account the uncertainties of each star}. We calculate the uncertainty in [X/Fe] by taking the quadrature of the uncertainties in [X/H] and [Fe/H]. 

To calculate the [X/Fe] and [X/H] cluster mean for each element, we generate $100\,000$ mock data points for each observed star, for which the abundances are drawn randomly from a Gaussian distribution centred at the observed abundances with standard deviation to be the uncertainty of the observed values. The median of the combined distribution is taken to be the cluster mean and $1\sigma$ of the combined distribution is estimated to be the uncertainty. The cluster mean determined in this way almost always coincides with the arithmetic mean. Note that if we were to use rms of the star-to-star dispersion to be $1\sigma$ of the cluster mean, the uncertainty would be much smaller ($\sigma \leq 0.03$).

\section[]{Results and discussion}\label{sec:results_and_discussion}

The derived abundances and total uncertainty per star per element are listed in Tables~\ref{table:EW_analysis_1}--\ref{table:EW_analysis_3}. We plot our results in Fig.~\ref{fig:xfe_feh}, and overplot field stars from \citet{red03,red06}, \citet{ben03} and recent open cluster compilation from \citet{car11} for comparison. For the open cluster compilation, we have 78 clusters with $6.4 \leq R_G \leq 20.8$ kpc. For each cluster, we take the mean abundance of the cluster for each element. 

\begin{table*}
\caption{Comparison with previous studies on IC 4756 giants.\label{table:previous_studies}}
\begin{tabular}{lccccccccccccccc}
\hline
Source           & Type      & Resolution & [Fe\,{\sc i}/H]&   [Na/Fe]      &    [Al/Fe]       &    [Mg/Fe]        &    [Si/Fe]      \\
\hline
This study       & 12 giants &  30\,000   & $-0.01\pm0.10$ & $0.21\pm0.15$  & $ 0.12\pm0.12$   & $ 0.12\pm0.13$    & $ 0.13\pm0.13$   \\
\citet{pac10}    & 3 giants  & 100\,000   & $ 0.08\pm0.11$ & $0.11\pm0.12$  & $-0.12\pm0.12$   & ---               & $ 0.02\pm0.11$   \\
\citet{smi09}    & 5 giants  &  48\,000   & $ 0.05\pm0.11$ & $0.02$         & ---              & $-0.05$           & $ 0.05\pm0.08$   \\
\citet{jac07}    & 6 giants  &  15\,000   & $-0.15\pm0.18$ & $0.57\pm0.19^a$& $ 0.29\pm0.21$   & ---               & $ 0.34\pm0.25$   \\
\citet{luc94}    & 3 giants  &  18\,000   & $-0.05\pm0.21$ & $0.14\pm0.22$  & $ 0.05\pm0.28^a$ & $ 0.22\pm0.40^a$  & $ 0.23\pm0.29$   \\
\vspace{-0.15cm} \\
Source           &    [Ca/Fe]     & [Ti\,{\sc i}/Fe] & [Cr\,{\sc i}/Fe] & [Cr\,{\sc ii}/Fe] &    [Ni/Fe]     &    [Zn/Fe]     &    [Ba/Fe]     \\
\vspace{-0.15cm} \\ 
This study       & $ 0.05\pm0.16$ & $0.00\pm0.19$    & $0.03\pm0.15$    & $0.25\pm0.14$     & $ 0.03\pm0.13$ & $0.06\pm0.13$  & $ 0.00\pm0.14$ \\
\citet{pac10}    & $-0.02\pm0.12$ & $0.03\pm0.13$    & $0.01\pm0.13$    & ---               & $-0.04\pm0.12$ & ---            & ---            \\
\citet{smi09}    & $ 0.02\pm0.09$ & $-0.05\pm0.09$   & $0.04\pm0.11$    & $0.19\pm0.11$     & $-0.01\pm0.06$ & ---            & ---            \\ 
\citet{jac07}    & $ 0.07\pm0.24$ & ---              & ---              & ---               & $ 0.08\pm0.21$ & ---            & ---            \\
\citet{luc94}    & $-0.07\pm0.35$ & $-0.08\pm0.31$   & $0.08\pm0.26$    & ---               & $ 0.08\pm0.30$ & ---            & $ 0.02\pm0.26$ \\
\hline
\end{tabular}
\flushleft
$^a$[X/H] does not have uncertainty estimation; we have no choice but to assume that all uncertainties come from [Fe/H]. The stated uncertainty is therefore smaller than it should be.
\end{table*}

\subsection[]{Light odd-Z elements: Na and Al}\label{sec:results_and_discussion}

For [Al/Fe], all stars from IC 4756 show homogeneous abundances. On the other hand, [Na/Fe] shows more intracluster star-to-star dispersion. As shown in Fig.~\ref{fig:teff}, abundances of Na are strongly dependent on the effective temperature with the two coolest stars (Her 6 with [Na/H] = 0.13 dex and Her 90 with [Na/H] = 0.04 dex) being the reason for the trend. As discussed in Section~\ref{subsection:ew}, the inhomogeneity may be explained by NLTE effects. This could, however, also be a sign of internal mixing in giants. The larger abundance scatter in Na is also consistent with the open cluster compilation values. As shown in Fig.~\ref{fig:xfe_feh}, among light odd-Z elements, [Na/Fe] is subjected to more scatter, presumably due to the choice of dwarfs or giants in each survey. This enhanced Na is also seen in field star surveys of clump and red giant stars \citep{siv09}, but this is not the case for [Al/Fe]. This observation is also consistent with field stars results from \citet{and08} and \citet{bon09} where they argued that Al does not suffer from significant internal mixing.

\subsection[]{$\alpha$-elements: Mg, Si, Ca and Ti}\label{subsection:alpha_elements}

Our analysis shows that [Mg/Fe] and [Si/Fe] are supersolar for IC 4756, but [Ca/Fe] and [Ti/Fe] seem to be solar. This is not new. Several high-resolution analyses of various open clusters show overabundances in lighter $\alpha$-elements \citep*[e.g.][]{yon05}. \citet{ting12} also showed that, compared to field stars, open cluster compilation shows more variation in $\alpha$-elements and more independent dimensions. This is consistent with the more interclusters scatter in lighter $\alpha$-elements, especially Mg, as shown in Fig.~\ref{fig:xfe_feh}. However, since Mg is usually derived with very few lines, the results should be viewed with caution.

We chose to use neutral Ti for our analysis. It is important to note that there is a discrepancy between Ti\,{\sc i} and Ti\,{\sc ii} abundances. Similar effects have previously been reported for stars cooler than \teff$<5200~K$ \citep{sch03}. The mechanism responsible for this, known as overionization effect, is the pumping of the electrons into the ionized state due to strong UV flux of the hot chromospheres \citep[refer to][for a more detailed discussion on the topic]{sch03,dor09}. Therefore we adopt the Ti\,{\sc i} abundance as a more reliable value.

\subsection[]{Fe-peak elements: Cr, Ni and Zn}\label{subsection:Fe_peak_elements}

All our Fe-peak elements are homogeneous, solar-like and show little intracluster star-to-star dispersion. As observed with Ti\,{\sc ii} abundances, we see the effects of overionization in Cr\,{\sc ii} as well. Therefore, we adopt the Cr\,{\sc i} abundance, which is in the solar proportions, in agreement with the other Fe-peak elements.

\subsection[]{Neutron-capture elements: Ba}\label{subsection:Neutron-capture}

Our study shows that Ba abundance is homogeneous and solar-like. This is strikingly different from what has been observed for the similar aged Hyades open cluster, where Ba is enhanced over 0.2 dex \citep{siv06,siv11}. It is often thought that different analysis methods and line lists might contribute to the scatter in [Ba/Fe]-[Fe/H], as shown in Fig.~\ref{fig:xfe_feh}. However our analysis adopts a very similar method compared to the studies of \citet{siv06,siv11}. It is therefore possible that the cluster-to-cluster dispersion (and star-to-star dispersion for field stars) in Ba abundance is real. This supports the idea that the metallicity-dependence nature of $s$-process in AGB phase \citep{bus01,cri09,bis10} creates a unique neutron-capture elements imprint on the gas from which the stars form \citep{ting12}.

\subsection[]{Comparison with previous studies}\label{subsec:previous_studies}

Her 85, Her 87 and Her 176 were previously studied by \citet{luc94}. Similar to our analysis, Her 85 also showed some elemental abundances that do not fit into the cluster mean. As discussed in Section~\ref{subsec:observation}, Her 85 is likely a non-member of IC 4756; therefore, we exclude Her 85 to calculate the cluster mean. For Her 87, they derived  \teff~$=5100$, $\log g= 2.8$, $v_t=2.40$, and for Her 176, they deduced \teff$=5200$, $\log g=3.0$, $v_t=1.90$. Furthermore, \citet{gil89} studied Her 87, Her 144, Her 176, Her 228, Her 249, Her 296 and Her 314. Other than some discrepancies in microturbulence, the comparison with our atmospheric parameters is very satisfactory. 

The comparison of elemental abundances with previous studies is listed in Table~\ref{table:previous_studies}. Note that, for a better comparison, we do not directly adopt the cluster mean and its uncertainty quoted in each study. {\it Instead, we calculate the cluster mean with the same method as described in Section~\ref{subsec:mean_abundances}}. 

All previous studies agree on [Fe/H], [Ca/Fe], [Ti/Fe], [Cr/Fe], [Ni/Fe] and [Ba/Fe], including [Fe/H] $= 0.04\pm0.20$ from seven giants of \citet{gil89} and [Fe/H] $=0.03\pm0.07$ from three giants of \citet{san09} that are not listed in Table~\ref{table:previous_studies}. All these abundances appear to be solar. Furthermore, [Cr\,{\sc ii}/Fe] is consistently about $0.2$ dex enhanced compared to [Cr\,{\sc i}/Fe], which mimics our results as well.

Our results are consistent with other studies that [Na/Fe] is above solar. Our analysis shows Na abundance in between the results of \citet{jac07} and \citet{pac10}. \citet{jac07} claimed that their Na abundance can be brought down if different line lists were considered and therefore might be more consistent with ours and \citet{pac10} results. However, \citet{smi09} claimed that there is no enhancement in Na if NLTE corrections is taken. 

On the other hand, there are discrepancies in term of [Al/Fe], [Mg/Fe] and [Si/Fe]. Our study agrees with \citet{jac07} that both Al and lighter $\alpha$-elements, such as Mg and Si, are enhanced. \citet{luc94} also reported the same for Mg and Si. \citet{pac10} reported a [Na/Fe] enhancement of 0.11~dex, which is similar to the level of enhancement in this study for [Al/Fe], [Mg/Fe] and [Si/Fe].

\section[]{Conclusions}\label{sec:conclusion}

We have performed a high-resolution spectroscopic abundance analysis on the intermediate-aged open cluster IC 4756, targeting abundances of Na, Al, Mg, Si, Ca, Ti, Cr, Ni, Fe, Zn and Ba for 12 subgiant cluster members. We find the cluster metallicity of [Fe/H] $=-0.01$ dex with abundance enhancements seen in Na, Al, Mg and Si, while all other elements are at solar level. Our study expands the total stars studied for this cluster and generally agrees well with the other literature work on this cluster.

We recover a high level of chemical homogeneity across a range of elements including odd-Z, $\alpha$, Fe-peak and neutron-capture elements up to Ba. This supports the concept that stars form from well-mixed gas clouds and that the chemical imprint of the protocluster cloud is preserved within the stars. This result of high chemical homogeneity further confirms that simple stellar populations such as open clusters should be chemically homogeneous. We therefore further validate the basic requirement for chemical tagging, namely other such stellar populations which are now dispersed into the Galaxy field could be identified via chemical tracers.

As an exception, we find a larger scatter in the Na abundance. The source of this scatter is unclear although enhanced Na abundances have been observed in this cluster by others. Also the Ba abundance was found to be at the solar level, which is vastly different from the enhanced Ba abundances observed in the Hyades. Given the similar age of the two clusters and the similar abundance analysis carried out in both clusters, we find this points to a real cosmic scatter in Ba among the open cluster population.

\section*{Acknowledgments}
YST is grateful to the Australian Astronomical Observatory, the College of Physical and Mathematical Sciences and the Research School of Astronomy and Astrophysics at The Australian National University for their financial support throughout this project. GMDS thanks the Apache Point Observatory for the observation time which was offered directly after the 2003 bushfire at Mt Stromlo Observatory.

\appendix

\section[]{Equivalent widths}\label{appendix:equivalent_widths}

\begin{table*}
\begin{center}
\caption{IC 4756 Na, Mg, Al, Si and Ca equivalent widths \label{table:EW_list_1}}
\begin{turn}{90}
\begin{tabular}{lccccccccccccccc}
\hline 
             &          &   &         & Her6  & Her35 & Her82 & Her87 & Her90 & Her144& Her170& Her176& Her228& Her249& Her296& Her397\\
             & $\lambda$&   &         & EW    & EW    & EW    & EW    & EW    & EW    & EW    & EW    & EW    & EW    & EW    & EW    \\
Element      & (\AA)    & EP&$\log gf$& (m\AA)& (m\AA)& (m\AA)& (m\AA)& (m\AA)& (m\AA)& (m\AA)& (m\AA)& (m\AA)& (m\AA)& (m\AA)& (m\AA)\\
\hline
Na\,{\sc i}  & 5682.63 & 2.10 & -0.71 & 141.6 & 132.9 & 127.7 & 127.5 & 143.6 & 141.9 & 122.6 & 119.9 & 133.8 & 129.6 & 127.3 & 135.3 \\
Na\,{\sc i}  & 5688.21 & 2.10 & -0.40 & 168.2 & 158.8 & 152.1 & 147.9 & 170.3 & 160.2 & 145.5 & 145.4 & 152.2 & 152.5 & 143.4 & 152.7 \\
Na\,{\sc i}  & 6154.23 & 2.10 & -1.57 &  86.2 &  75.4 &  71.9 &  73.6 &  87.7 &  82.3 &  71.3 &  69.6 &  73.8 &  77.1 &  70.1 &  77.9 \\
Na\,{\sc i}  & 6160.75 & 2.10 & -1.27 & 107.0 &  93.7 &  92.8 &  93.8 & 107.8 & 104.2 &  90.0 &  86.2 &  94.8 &  95.6 &  88.1 &  98.4 \\
\\
Mg\,{\sc i}  & 5711.09 & 4.35 & -1.86 & 135.1 & 121.5 & 122.7 & 122.9 & 138.5 & 127.9 & 117.5 & 113.9 & 123.4 & 120.6 & 112.0 & 122.3 \\
\\
Al\,{\sc i}  & 5557.06 & 3.14 & -2.21 &  34.5 &  25.7 &  23.7 &  26.5 &  ---  &  27.0 &  23.4 &  24.2 &  24.6 &  26.0 &  23.6 &  25.4 \\
Al\,{\sc i}  & 6698.67 & 3.14 & -1.92 &  53.0 &  41.2 &  42.9 &  43.3 &  57.2 &  41.8 &  39.0 &  42.7 &  36.2 &  40.3 &  37.8 &  43.3 \\ 
\\
Si\,{\sc i}  & 5645.61 & 4.93 & -2.04 & ---   &  59.6 &  59.5 & ---   & ---   &  59.7 &  52.4 &  54.6 &  59.0 &  63.2 &  56.8 &  64.4 \\
Si\,{\sc i}  & 5665.56 & 4.92 & -1.94 &  67.5 &  69.7 &  59.1 &  60.9 & ---   &  69.0 &  59.4 &  61.9 &  70.1 &  64.4 &  61.3 &  68.3 \\
Si\,{\sc i}  & 5684.48 & 4.95 & -1.55 &  80.5 &  92.0 &  82.9 &  87.0 &  81.1 &  87.8 &  75.8 &  81.0 &  87.0 &  77.6 &  79.6 &  89.3 \\
Si\,{\sc i}  & 5690.42 & 4.93 & -1.77 &  69.5 &  77.3 &  69.7 &  74.9 &  73.8 &  67.5 &  63.2 &  63.9 &  74.1 &  73.3 &  66.5 &  71.4 \\
Si\,{\sc i}  & 5701.10 & 4.93 & -1.95 &  60.5 &  63.8 &  63.4 &  65.0 &  54.8 &  60.3 &  54.8 &  54.6 &  64.7 &  60.2 &  61.1 &  60.6 \\
Si\,{\sc i}  & 5948.54 & 5.08 & -1.13 & 102.9 & 103.5 & 103.2 & 100.9 & 102.3 & 102.1 &  95.1 &  91.8 & 105.2 &  96.7 &  98.2 & 100.0 \\
Si\,{\sc i}  & 6125.02 & 5.61 & -1.52 &  43.0 &  44.8 &  49.2 &  51.9 &  42.3 &  48.3 &  49.0 &  42.8 &  49.9 &  48.5 &  42.6 &  49.0 \\
Si\,{\sc i}  & 6142.48 & 5.62 & -1.50 &  47.9 &  43.2 &  53.4 &  51.6 &  45.0 &  41.0 &  40.5 &  47.2 &  48.0 &  46.0 &  49.4 &  53.4 \\
Si\,{\sc i}  & 6145.02 & 5.62 & -1.45 &  51.3 &  57.9 &  49.1 &  54.1 &  39.3 &  44.7 &  45.3 &  44.3 &  51.6 &  45.4 &  47.3 &  50.9 \\
Si\,{\sc i}  & 6155.13 & 5.62 & -0.72 &  96.3 & 101.2 &  91.7 &  95.2 &  83.6 &  91.7 &  85.1 &  90.9 &  98.3 &  90.1 &  88.5 &  95.5 \\
Si\,{\sc i}  & 6237.32 & 5.61 & -1.05 &  74.7 &  77.0 &  81.4 &  78.3 &  69.3 &  72.3 &  70.2 &  71.2 &  73.1 &  80.6 &  76.6 &  79.2 \\
Si\,{\sc i}  & 6243.81 & 5.62 & -1.29 & ---   &  63.6 &  61.0 &  63.4 &  53.0 &  57.0 &  60.9 &  51.2 &  63.2 &  66.7 &  57.7 &  65.4 \\
Si\,{\sc i}  & 6244.47 & 5.62 & -1.32 &  51.3 &  63.2 &  59.9 &  59.5 &  45.4 &  54.6 &  56.9 &  58.2 &  65.9 &  64.0 &  55.2 &  63.9 \\
\\
Ca\,{\sc i}  & 5260.39 & 2.52 & -1.78 &  78.5 &  64.9 &  60.7 &  63.4 & ---   &  62.5 &  52.4 &  55.4 &  58.5 &  61.0 &  57.3 &  62.6 \\
Ca\,{\sc i}  & 5261.70 & 2.52 & -0.45 & 156.6 & 134.2 & 127.4 & 124.9 & ---   & 135.4 & 121.9 & 126.3 & 136.3 & 133.6 & 126.7 & 137.6 \\ 
Ca\,{\sc i}  & 5349.46 & 2.71 & -0.64 & 126.3 & 120.6 & 111.7 & 110.2 & ---   & 121.3 & 102.9 & 101.4 & 114.9 & 106.3 & 108.4 & 108.2 \\
Ca\,{\sc i}  & 5512.98 & 2.93 & -0.56 & 119.1 & 108.7 & 102.6 & 106.3 & 128.9 & 110.2 &  92.6 &  94.5 & 105.8 & 108.9 &  93.1 & 104.3 \\
Ca\,{\sc i}  & 5590.11 & 2.52 & -0.52 & ---   & 137.3 & 122.0 & 123.9 & ---   & 133.8 & 120.2 & 118.1 & 131.0 & 125.3 & 119.2 & 133.6 \\
Ca\,{\sc i}  & 5601.28 & 2.53 & -0.25 & 171.9 & 157.0 & 145.7 & 148.3 & 204.4 & 156.8 & 130.4 & 129.4 & 146.9 & 144.3 & 134.2 & 145.7 \\
Ca\,{\sc i}  & 5867.56 & 2.93 & -1.60 &  62.4 &  47.8 &  42.1 &  51.8 &  63.9 &  43.9 &  42.1 &  44.3 &  53.4 &  46.6 &  43.6 &  48.7 \\
Ca\,{\sc i}  & 6122.22 & 1.89 & -0.37 & ---   & 213.7 & 201.4 & 203.6 & 263.0 & 204.8 & 189.6 & 194.8 & 209.4 & 200.9 & 189.2 & 202.2 \\
Ca\,{\sc i}  & 6161.30 & 2.52 & -1.26 & 114.8 &  95.8 &  89.0 &  84.5 & ---   & 101.3 &  79.6 &  80.4 &  94.5 &  89.9 &  84.2 &  95.7 \\
Ca\,{\sc i}  & 6166.44 & 2.52 & -1.17 & 119.2 & 101.2 &  96.6 & 100.5 & 124.4 & 108.6 &  95.9 &  83.2 &  98.2 &  95.5 &  94.3 & 100.6 \\
Ca\,{\sc i}  & 6169.04 & 2.52 & -0.84 & 135.4 & 123.9 & 110.3 & 109.0 & 136.8 & 128.4 & 107.2 & 105.6 & 110.5 & 107.5 & 101.0 & 119.3 \\
Ca\,{\sc i}  & 6169.56 & 2.53 & -0.62 & 145.3 & 138.6 & 121.0 & 123.2 & 167.0 & 131.7 & 123.7 & 113.3 & 125.5 & 119.0 & 114.9 & 123.5 \\
Ca\,{\sc i}  & 6455.60 & 2.52 & -1.30 & 104.1 &  95.3 &  86.2 &  86.9 & 125.5 & 103.2 &  87.5 &  77.0 &  94.4 &  92.7 &  84.9 &  88.0 \\
Ca\,{\sc i}  & 6471.66 & 2.53 & -0.64 & 143.7 & 132.2 & 122.4 & 127.7 & 157.6 & 135.8 & 119.7 & 115.0 & 132.6 & 122.2 & 112.9 & 132.9 \\
Ca\,{\sc i}  & 6493.78 & 2.52 & -0.19 & 186.0 & 193.6 & 160.6 & 162.4 & 208.1 & 172.1 & 144.4 & 142.5 & 165.2 & 166.3 & 155.3 & 171.9 \\
Ca\,{\sc i}  & 6499.65 & 2.52 & -0.73 & 137.3 & 122.7 & 118.2 & 120.7 & 152.0 & ---   & 110.6 & 109.3 & 126.2 & 122.1 & 112.9 & 121.6 \\
Ca\,{\sc i}  & 6798.48 & 2.71 & -2.46 &  25.5 &  20.2 &  20.2 & ---   &  38.6 &  20.4 & ---   &  20.8 &  20.4 &  21.4 &  21.5 &  19.6 \\
\hline
\end{tabular}
\end{turn}
\end{center}
\end{table*}

\begin{table*}
\begin{center}
\caption{IC 4756 Ti and Cr equivalent widths \label{table:EW_list_2}}
\begin{turn}{90}
\begin{tabular}{lccccccccccccccc}
\hline
             &          &   &         & Her6  & Her35 & Her82 & Her87 & Her90 & Her144& Her170& Her176& Her228& Her249& Her296& Her397\\
             & $\lambda$&   &         & EW    & EW    & EW    & EW    & EW    & EW    & EW    & EW    & EW    & EW    & EW    & EW    \\
Element      & (\AA)    & EP&$\log gf$& (m\AA)& (m\AA)& (m\AA)& (m\AA)& (m\AA)& (m\AA)& (m\AA)& (m\AA)& (m\AA)& (m\AA)& (m\AA)& (m\AA)\\
\hline
Ti\,{\sc i}  & 4840.87 & 0.90 & -0.45 & ---   & 112.7 & 104.4 & 109.1 & 149.8 & 119.4 &  98.5 &  99.8 & 111.5 & 106.3 &  97.9 & 107.3 \\
Ti\,{\sc i}  & 4913.61 & 1.87 &  0.22 & ---   &  94.8 &  85.2 & ---   & 124.7 & ---   &  82.1 &  79.7 &  87.2 &  86.8 &  81.4 &  87.8 \\
Ti\,{\sc i}  & 5016.16 & 0.85 & -0.52 & 140.1 & 114.3 & 107.7 & 109.4 & 149.9 & 124.1 & 101.3 & 104.0 & 113.3 & 107.3 &  99.0 & 102.5 \\
Ti\,{\sc i}  & 5022.87 & 0.83 & -0.38 & 153.4 & 120.8 & 116.3 & 114.4 & 165.2 & 125.8 & 112.1 & 106.9 & 114.1 & 121.1 & 106.5 & 117.6 \\ 
Ti\,{\sc i}  & 5024.84 & 0.82 & -0.55 & 139.1 & 121.3 & 112.0 & 111.9 & 154.0 & 124.5 & 102.6 &  98.1 & 112.5 & 110.8 & 102.9 & 111.2 \\
Ti\,{\sc i}  & 5087.06 & 1.43 & -0.78 &  98.5 & ---   &  72.6 &  68.6 & ---   &  78.4 &  68.5 &  66.4 &  69.7 &  68.8 &  67.7 &  72.3 \\
Ti\,{\sc i}  & 5113.44 & 1.44 & -0.73 &  93.2 &  67.2 &  67.0 &  72.4 & ---   &  76.8 &  62.9 &  70.6 &  75.6 &  72.2 &  63.6 &  70.5 \\
Ti\,{\sc i}  & 5219.70 & 0.02 & -2.24 & 111.2 &  84.4 &  79.5 & ---   & 122.9 &  91.8 &  69.3 &  74.6 &  77.5 &  75.3 &  71.4 &  74.8 \\
Ti\,{\sc i}  & 5426.25 & 0.02 & -2.95 &  75.0 &  36.1 &  40.3 & ---   &  87.4 &  44.7 &  38.9 &  39.2 &  43.0 &  38.6 &  32.3 &  40.1 \\
Ti\,{\sc i}  & 5471.19 & 1.44 & -1.61 &  56.5 &  30.8 &  29.1 &  31.8 &  65.6 &  28.8 &  26.4 &  28.0 &  26.7 &  32.0 &  24.9 &  27.5 \\
Ti\,{\sc i}  & 5490.15 & 1.46 & -0.88 &  93.7 &  66.6 &  59.3 &  62.6 & 100.9 &  67.9 &  62.6 &  60.1 &  60.0 &  67.7 &  59.0 &  56.3 \\
Ti\,{\sc i}  & 5866.45 & 1.07 & -0.78 & 129.4 & 104.2 &  97.0 &  94.4 & 136.5 & 107.7 &  90.0 &  89.3 &  93.5 &  99.2 &  85.5 &  94.8 \\ 
Ti\,{\sc i}  & 5953.16 & 1.89 & -0.27 & 102.9 &  68.9 &  72.1 &  77.7 & 109.7 &  78.1 &  71.8 & ---   &  75.9 & ---   &  71.2 &  71.1 \\
Ti\,{\sc i}  & 6126.22 & 1.07 & -1.37 & 100.7 &  63.3 &  60.8 &  67.6 & 111.7 &  74.5 &  63.6 &  62.7 &  69.1 &  66.1 &  62.8 &  64.6 \\
Ti\,{\sc i}  & 6258.10 & 1.44 & -0.30 & ---   &  99.2 &  93.5 & ---   & 140.2 & ---   & ---   &  93.3 &  94.7 &  95.6 &  87.0 &  96.6 \\
Ti\,{\sc i}  & 6261.10 & 1.43 & -0.42 & ---   &  95.0 &  94.1 &  97.2 & ---   & 105.3 &  87.2 &  87.3 &  97.5 &  99.0 &  89.7 &  93.0 \\
Ti\,{\sc i}  & 6303.76 & 1.44 & -1.51 &  65.8 &  29.7 &  40.5 &  40.8 &  76.5 &  35.4 &  33.4 &  33.0 &  33.0 &  35.7 &  31.7 &  31.1 \\
Ti\,{\sc i}  & 6743.12 & 0.90 & -1.63 & ---   &  64.0 &  63.6 &  71.5 & 104.9 &  71.6 &  62.8 &  66.0 &  69.5 &  63.0 &  58.6 &  55.7 \\
\\
Ti\,{\sc ii} & 4865.61 & 1.12 & -2.79 &  86.0 &  79.5 &  77.4 &  80.5 & ---   &  89.5 &  83.7 &  74.4 &  82.7 &  78.0 &  77.4 &  81.0 \\
Ti\,{\sc ii} & 4911.19 & 3.12 & -0.61 &  78.2 & ---   &  80.1 &  79.9 & ---   &  89.0 &  81.6 & ---   &  82.0 &  77.8 &  76.5 &  81.2 \\
Ti\,{\sc ii} & 5005.16 & 1.57 & -2.72 &  61.2 &  64.5 &  57.7 &  58.3 &  66.5 &  67.4 &  61.2 & ---   & ---   & ---   &  55.7 &  53.3 \\
Ti\,{\sc ii} & 5154.07 & 1.57 & -1.75 & ---   & 112.7 & 107.2 & 103.2 & ---   & 113.1 & 105.2 &  99.6 & 108.0 & 103.6 & 102.5 & 104.0 \\
Ti\,{\sc ii} & 5185.91 & 1.89 & -1.49 & 106.3 & 102.6 &  97.2 &  99.4 & 105.7 & 107.7 & 100.6 &  96.0 & 105.4 & 100.8 &  98.6 & 102.1 \\
Ti\,{\sc ii} & 5336.77 & 1.58 & -1.59 & 115.9 & 117.1 & 115.7 & 112.5 & 121.6 & 123.5 & 109.9 & 108.0 & 116.0 & 107.0 & 106.0 & 111.6 \\
Ti\,{\sc ii} & 5381.02 & 1.57 & -1.92 & 101.8 & 105.5 &  93.3 &  95.3 & 108.7 & 106.5 &  99.4 &  88.9 &  98.9 &  97.2 &  92.9 &  96.7 \\
Ti\,{\sc ii} & 5490.69 & 1.57 & -2.43 &  81.1 &  76.9 &  72.2 &  76.3 & ---   &  82.1 &  74.4 &  69.6 &  76.9 &  74.3 &  73.2 &  75.2 \\
\\
Cr\,{\sc i}  & 5238.96 & 2.71 & -1.43 &  52.9 &  41.4 &  35.6 &  40.9 &  64.3 &  39.8 &  35.5 &  35.4 & ---   &  38.6 &  32.2 &  39.8 \\
Cr\,{\sc i}  & 5296.69 & 0.98 & -1.32 & 170.1 & 143.6 & 136.2 & 135.6 & 196.8 & ---   & 127.5 & 124.3 & 138.5 & 137.6 & 126.7 & 138.9 \\
Cr\,{\sc i}  & 5304.18 & 3.46 & -0.77 &  45.0 &  33.6 &  28.1 &  34.0 &  58.4 &  32.8 &  29.1 &  28.1 &  32.4 &  31.8 &  27.7 &  31.3 \\
Cr\,{\sc i}  & 5318.81 & 3.44 & -0.77 &  43.2 &  29.4 &  30.1 &  34.5 & ---   &  36.9 &  29.5 &  32.4 &  32.0 &  35.1 &  33.7 &  37.1 \\
Cr\,{\sc i}  & 6330.09 & 0.94 & -2.90 & 104.1 &  78.8 &  67.1 &  72.2 & 117.0 &  77.5 &  70.1 &  68.1 &  72.1 &  74.7 &  68.0 &  69.9 \\
\\
Cr\,{\sc ii} & 5237.33 & 4.07 & -1.18 &  67.6 &  81.0 &  74.9 &  76.1 &  72.9 &  80.1 &  74.4 &  70.4 &  75.8 &  75.2 &  72.8 &  73.0 \\
Cr\,{\sc ii} & 5305.85 & 3.83 & -2.06 &  42.8 &  51.0 &  46.6 &  47.9 &  38.3 &  49.8 &  45.8 &  45.2 &  47.6 &  47.5 &  45.8 &  46.5 \\
Cr\,{\sc ii} & 5308.41 & 4.07 & -1.79 &  46.2 &  52.9 &  48.4 &  47.0 &  39.6 &  48.6 &  46.9 &  49.7 &  46.8 &  47.2 &  47.7 &  49.6 \\
Cr\,{\sc ii} & 5310.69 & 4.07 & -2.24 & ---   &  28.8 &  29.0 &  30.2 & ---   &  28.2 &  32.2 &  28.7 &  30.9 &  30.5 &  28.8 &  30.4 \\
Cr\,{\sc ii} & 5313.56 & 4.07 & -1.55 &  54.6 &  62.7 &  54.7 &  60.7 &  52.4 &  61.1 &  57.0 &  56.9 &  59.5 &  57.4 &  56.6 &  60.9 \\
\hline
\end{tabular}
\end{turn}
\end{center}
\end{table*}

\begin{table*}
\begin{center}
\caption{IC 4756 Fe equivalent widths \label{table:EW_list_3}}
\begin{turn}{90}
\begin{tabular}{lccccccccccccccc}
\hline
             &          &   &         & Her6  & Her35 & Her82 & Her87 & Her90 & Her144& Her170& Her176& Her228& Her249& Her296& Her397\\
             & $\lambda$&   &         & EW    & EW    & EW    & EW    & EW    & EW    & EW    & EW    & EW    & EW    & EW    & EW    \\
Element      & (\AA)    & EP&$\log gf$& (m\AA)& (m\AA)& (m\AA)& (m\AA)& (m\AA)& (m\AA)& (m\AA)& (m\AA)& (m\AA)& (m\AA)& (m\AA)& (m\AA)\\
\hline
Fe\,{\sc i}  & 4347.24 & 0.00 & -5.50 & ---   & ---   &  83.6 & ---   & ---   & ---   & ---   & ---   & ---   & ---   & ---   & ---   \\
Fe\,{\sc i}  & 4439.88 & 2.28 & -3.00 & ---   & ---   & ---   & ---   & ---   &  94.5 & ---   & ---   & ---   & ---   & ---   & ---   \\
Fe\,{\sc i}  & 4442.34 & 2.20 & -1.26 & 231.2 & ---   & 193.4 & 227.0 & ---   & ---   & 200.0 & 183.1 & ---   & 218.0 & 195.6 & ---   \\
Fe\,{\sc i}  & 4445.48 & 0.08 & -5.44 & ---   & ---   &  84.5 & ---   & 112.9 &  95.0 &  81.0 & ---   &  84.2 & ---   &  76.3 & ---   \\
Fe\,{\sc i}  & 4447.72 & 2.22 & -1.34 & 220.6 & ---   & 185.4 & 203.1 & ---   & ---   & 183.4 & 175.4 & 195.9 & 193.3 & 174.6 & ---   \\
Fe\,{\sc i}  & 4494.57 & 2.20 & -1.14 & ---   & ---   & ---   & 214.4 & ---   & ---   & 201.3 & ---   & ---   & ---   & ---   & ---   \\
Fe\,{\sc i}  & 4523.40 & 3.65 & -1.99 & ---   & ---   & ---   & ---   &  80.8 & ---   & ---   & ---   & ---   &  60.6 & ---   & ---   \\
Fe\,{\sc i}  & 4547.85 & 3.54 & -1.01 & ---   & ---   & ---   & ---   & ---   & 120.5 & 104.5 & ---   & ---   & ---   & 104.1 & ---   \\
Fe\,{\sc i}  & 4556.93 & 3.25 & -2.71 & ---   &  57.7 & ---   &  57.9 & ---   & ---   &  49.1 & ---   & ---   & ---   &  49.3 &  48.2 \\
Fe\,{\sc i}  & 4561.43 & 2.76 & -3.08 & ---   & ---   & ---   & ---   & ---   &  65.0 & ---   & ---   & ---   & ---   & ---   & ---   \\ 
Fe\,{\sc i}  & 4593.52 & 3.94 & -2.06 &  51.5 &  48.0 & ---   &  44.3 & ---   & ---   & ---   & ---   &  42.8 & ---   &  41.8 &  40.4 \\
Fe\,{\sc i}  & 4602.01 & 1.61 & -3.15 & 130.6 & ---   & 102.6 & ---   & 139.7 & ---   & 100.9 & 100.0 & 112.2 & ---   &  99.9 & ---   \\
Fe\,{\sc i}  & 4635.84 & 2.85 & -2.36 & ---   & ---   & ---   & ---   & ---   &  98.0 & ---   &  80.6 & ---   & ---   & ---   & ---   \\
Fe\,{\sc i}  & 4678.84 & 3.60 & -0.83 & ---   & ---   & ---   & ---   & ---   & 130.5 & ---   & ---   & ---   & ---   & ---   & ---   \\
Fe\,{\sc i}  & 4679.22 & 3.25 & -2.42 & ---   & ---   & ---   &  68.8 & ---   & ---   & ---   & ---   & ---   & ---   &  60.8 & ---   \\
Fe\,{\sc i}  & 4683.55 & 2.83 & -2.32 & ---   & ---   & ---   &  83.5 & 109.6 & ---   & ---   & ---   &  86.3 &  85.1 & ---   &  86.3 \\    
Fe\,{\sc i}  & 4704.94 & 3.69 & -1.53 & ---   & ---   & ---   &  83.6 & ---   &  85.5 & ---   & ---   & ---   & ---   & ---   & ---   \\
Fe\,{\sc i}  & 4705.45 & 3.55 & -2.27 &  71.6 &  65.0 &  55.6 &  65.4 &  74.9 & ---   &  56.7 &  52.6 & ---   & ---   & ---   &  57.6 \\
Fe\,{\sc i}  & 4779.43 & 3.42 & -2.02 & ---   & ---   &  71.2 &  71.0 & ---   &  79.0 & ---   & ---   & ---   &  74.7 &  67.7 & ---   \\ 
Fe\,{\sc i}  & 4787.82 & 3.00 & -2.60 & ---   &  68.0 &  63.4 &  70.7 & ---   &  81.5 &  66.8 &  62.2 &  72.2 & ---   & ---   & ---   \\
Fe\,{\sc i}  & 4788.75 & 3.24 & -1.76 & ---   & ---   &  90.1 &  97.2 & 114.0 & 105.5 &  90.7 &  85.4 & ---   &  98.0 &  86.3 &  92.1 \\
Fe\,{\sc i}  & 4802.87 & 3.64 & -1.51 & ---   & ---   & ---   & ---   & ---   & ---   & ---   & ---   &  85.1 & ---   & ---   & ---   \\
Fe\,{\sc i}  & 4809.93 & 3.57 & -2.68 &  46.7 &  38.2 &  40.3 &  41.3 & ---   &  40.0 &  40.8 &  34.9 &  37.4 &  39.0 &  36.0 &  29.8 \\
Fe\,{\sc i}  & 4869.46 & 3.55 & -2.48 &  60.5 &  53.3 &  48.8 &  56.3 & ---   &  54.0 &  48.3 & ---   &  53.9 &  53.6 &  44.1 &  50.0 \\  
Fe\,{\sc i}  & 4961.91 & 3.63 & -2.25 & ---   &  49.6 & ---   &  50.0 &  64.3 &  60.0 & ---   & ---   & ---   & ---   & ---   &  49.4 \\
Fe\,{\sc i}  & 5054.64 & 3.64 & -1.92 & ---   & ---   & ---   & ---   & ---   & ---   & ---   & ---   & ---   & ---   & ---   &  70.1 \\
Fe\,{\sc i}  & 5293.95 & 4.14 & -1.84 & ---   & ---   & ---   &  52.8 &  67.2 & ---   &  49.3 &  52.2 & ---   & ---   & ---   & ---   \\
Fe\,{\sc i}  & 5294.54 & 3.64 & -2.81 & ---   &  30.0 &  27.0 &  23.8 &  46.9 & ---   &  30.6 &  30.9 &  30.4 &  28.6 &  24.5 &  26.0 \\
Fe\,{\sc i}  & 5386.33 & 4.15 & -1.74 &  59.9 &  56.7 &  44.8 &  53.5 &  62.9 &  52.0 &  52.4 &  47.1 &  52.9 &  52.2 & ---   &  48.5 \\
Fe\,{\sc i}  & 5417.03 & 4.42 & -1.66 & ---   & ---   &  48.3 & ---   & ---   & ---   & ---   &  46.0 & ---   & ---   & ---   & ---   \\
Fe\,{\sc i}  & 5466.99 & 3.57 & -2.44 & ---   & ---   &  55.1 & ---   & ---   & ---   & ---   &  52.2 & ---   & ---   & ---   & ---   \\
Fe\,{\sc i}  & 5618.63 & 4.21 & -1.29 & ---   & ---   & ---   & ---   & ---   &  72.8 & ---   & ---   &  78.8 &  78.7 &  71.3 & ---   \\ 
Fe\,{\sc i}  & 5633.95 & 4.99 & -0.27 & ---   &  83.4 &  81.7 &  80.7 & ---   &  86.5 & ---   &  71.8 &  84.1 &  81.1 &  77.0 &  79.1 \\
Fe\,{\sc i}  & 5662.52 & 4.16 & -0.52 & ---   & ---   & ---   & ---   & ---   & ---   & ---   & 107.2 & ---   & ---   & ---   & ---   \\
Fe\,{\sc i}  & 5701.55 & 2.56 & -2.22 & ---   & ---   & ---   & 109.6 & ---   & ---   & ---   & ---   & ---   & ---   & ---   & ---   \\
Fe\,{\sc i}  & 5705.47 & 4.30 & -1.42 & ---   &  60.7 & ---   & ---   & ---   &  59.8 & ---   & ---   &  58.9 & ---   & ---   &  55.5 \\
Fe\,{\sc i}  & 5741.85 & 4.25 & -1.69 &  62.1 &  53.6 &  47.4 &  53.1 & ---   &  54.2 &  48.3 &  51.5 &  53.6 &  52.6 &  47.8 &  53.7 \\
Fe\,{\sc i}  & 5775.08 & 4.22 & -1.31 &  84.6 &  75.0 &  75.7 & ---   & ---   &  82.7 & ---   & ---   & ---   & ---   & ---   &  74.1 \\
\hline
\end{tabular}
\end{turn}
\end{center}
\end{table*}

\begin{table*}
\begin{center}
\begin{turn}{90}
\begin{tabular}{lccccccccccccccc}
\hline
             &          &   &         & Her6  & Her35 & Her82 & Her87 & Her90 & Her144& Her170& Her176& Her228& Her249& Her296& Her397\\
             & $\lambda$&   &         & EW    & EW    & EW    & EW    & EW    & EW    & EW    & EW    & EW    & EW    & EW    & EW    \\
Element      & (\AA)    & EP&$\log gf$& (m\AA)& (m\AA)& (m\AA)& (m\AA)& (m\AA)& (m\AA)& (m\AA)& (m\AA)& (m\AA)& (m\AA)& (m\AA)& (m\AA)\\
\hline
Fe\,{\sc i}  & 5778.48 & 2.59 & -3.48 &  64.2 &  50.5 &  54.0 &  57.5 &  78.0 &  55.0 &  49.6 &  51.4 &  58.0 &  50.4 &  47.8 &  53.9 \\
Fe\,{\sc i}  & 5811.92 & 4.14 & -2.43 &  23.5 &  20.7 & ---   & ---   &  29.1 &  20.4 &  24.4 & ---   &  24.2 &  26.6 &  19.6 &  18.9 \\
Fe\,{\sc i}  & 5837.70 & 4.29 & -2.34 & ---   &  19.8 & ---   & ---   &  23.4 & ---   & ---   & ---   & ---   & ---    & ---   & ---   \\
Fe\,{\sc i}  & 5853.16 & 1.49 & -5.28 &  47.4 &  27.1 &  25.6 &  31.4 &  48.4 &  29.6 &  27.7 & ---   &  26.2 &  24.5 &  21.8 &  22.3 \\
Fe\,{\sc i}  & 5855.09 & 4.60 & -1.55 &  40.4 &  40.4 &  39.4 &  38.5 &  54.2 &  41.3 &  39.0 &  36.8 &  42.0 &  41.5 &  37.1 &  35.3 \\
Fe\,{\sc i}  & 5856.10 & 4.29 & -1.64 &  60.2 & ---   & ---   &  59.8 & ---   &  58.7 & ---   & ---   & ---   & ---   & ---   & ---   \\
Fe\,{\sc i}  & 5858.75 & 4.22 & -2.26 &  28.4 &  22.3 &  24.2 &  23.5 &  32.5 &  23.7 & ---   &  24.5 &  24.7 &  24.3 & ---   &  22.2 \\ 
Fe\,{\sc i}  & 5927.80 & 4.65 & -1.09 &  62.5 &  58.9 &  54.6 &  57.5 &  67.7 &  62.3 &  54.9 & ---   &  58.4 &  58.8 &  54.0 &  57.3 \\
Fe\,{\sc i}  & 5956.69 & 0.86 & -4.61 & 124.2 & ---   & ---   & 103.4 & ---   & 104.1 &  97.6 & ---   & ---   &  98.2 &  92.5 &  91.5 \\
Fe\,{\sc i}  & 6015.24 & 2.22 & -4.76 & ---   &  18.9 & ---   &  16.4 & ---   &  17.8 &  17.4 &  15.9 & ---   &  20.3 & ---   & ---   \\
Fe\,{\sc i}  & 6024.06 & 4.55 &  0.16 & ---   & ---   & ---   & ---   & ---   & ---   & ---   & 127.2 & 133.1 & ---   & ---   & ---   \\ 
Fe\,{\sc i}  & 6034.04 & 4.31 & -2.47 & ---   & ---   & ---   & ---   & ---   & ---   & ---   & ---   & ---   &  21.0 &  16.4 & ---   \\
Fe\,{\sc i}  & 6042.23 & 4.65 & -0.89 &  82.5 &  75.0 &  70.2 &  74.7 &  82.7 &  81.4 & ---   &  72.5 &  75.5 &  75.7 &  69.1 & ---   \\
Fe\,{\sc i}  & 6054.08 & 4.37 & -2.31 &  19.3 &  14.2 & ---   & ---   &  21.0 & ---   & ---   &  22.8 &  16.0 & ---   &  15.6 &  16.4 \\   
Fe\,{\sc i}  & 6056.01 & 4.73 & -0.65 & ---   &  84.7 &  77.2 & ---   & ---   & ---   &  78.5 & ---   &  84.3 &  85.3 &  75.7 &  82.7 \\
Fe\,{\sc i}  & 6093.65 & 4.61 & -1.51 &  50.1 &  48.0 &  37.2 &  46.9 &  52.8 &  47.7 & ---   &  43.6 & ---   &  49.1 &  43.2 &  39.0 \\
Fe\,{\sc i}  & 6094.37 & 4.65 & -1.65 & ---   &  38.9 &  31.2 &  33.5 &  46.5 &  35.8 &  32.7 &  31.0 & ---   &  38.9 & ---   &  33.8 \\ 
Fe\,{\sc i}  & 6096.67 & 3.98 & -1.88 &  62.7 &  54.4 &  59.5 &  58.5 &  68.1 &  54.4 &  52.6 & ---   &  59.2 &  57.9 &  55.6 &  56.0 \\
Fe\,{\sc i}  & 6105.13 & 4.55 & -1.99 &  26.2 &  20.7 &  17.8 &  24.8 &  29.5 &  21.4 &  21.9 &  24.8 &  22.7 & ---   & ---   &  21.5 \\ 
Fe\,{\sc i}  & 6120.24 & 0.91 & -5.97 & ---   &  25.4 &  23.5 &  28.6 & ---   & ---   & ---   &  28.9 &  26.3 &  26.0 &  21.0 & ---   \\
Fe\,{\sc i}  & 6151.62 & 2.17 & -3.30 & ---   & ---   &  79.7 & ---   & 114.1 & ---   & ---   &  82.5 &  90.8 &  87.4 & ---   &  86.7 \\
Fe\,{\sc i}  & 6157.73 & 4.08 & -1.32 & ---   &  88.2 & ---   & ---   & ---   &  89.6 & ---   & ---   & ---   & ---   & ---   & ---   \\
Fe\,{\sc i}  & 6159.38 & 4.61 & -1.97 & ---   &  21.8 &  24.2 &  27.7 & ---   & ---   &  21.6 &  23.5 & ---   &  20.9 &  22.5 & ---   \\
Fe\,{\sc i}  & 6173.34 & 2.22 & -2.88 & 127.1 & 109.5 & 103.0 & 111.3 & 140.2 & ---   &  99.4 &  98.4 & 109.2 & 106.8 &  99.4 & ---   \\
Fe\,{\sc i}  & 6180.20 & 2.73 & -2.64 & ---   &  97.1 &  92.6 & ---   & ---   & 101.9 & ---   & ---   & ---   & ---   & ---   & ---   \\
Fe\,{\sc i}  & 6200.31 & 2.61 & -2.44 & ---   & 106.5 &  98.9 & 108.1 & 132.8 & 111.9 & 102.3 & 100.9 & 107.9 & 109.6 &  97.0 & ---   \\
\\ 
Fe\,{\sc ii} & 4491.41 & 2.86 & -2.68 & ---   & ---   & ---   & ---   & ---   & ---   &  95.1 & ---   & ---   &  89.3 & ---   & ---   \\
Fe\,{\sc ii} & 4508.28 & 2.86 & -2.31 & ---   & ---   & 104.6 & ---   & ---   & 116.8 & 105.7 &  99.7 & 108.2 & 105.1 & ---   & ---   \\
Fe\,{\sc ii} & 4541.52 & 2.84 & -2.99 & ---   & ---   & ---   & ---   & ---   & ---   &  83.7 & ---   & ---   & ---   & ---   & ---   \\
Fe\,{\sc ii} & 4576.33 & 2.84 & -2.82 & ---   & ---   & ---   & ---   & ---   & ---   &  85.4 & ---   & ---   & ---   & ---   & ---   \\
Fe\,{\sc ii} & 4582.83 & 2.84 & -3.09 &  74.1 &  77.2 & ---   & ---   & ---   &  83.8 &  77.3 & ---   &  81.6 &  75.6 & ---   & ---   \\ 
Fe\,{\sc ii} & 4620.52 & 2.83 & -3.08 &  71.1 & ---   &  76.0 &  75.1 &  72.9 &  87.5 &  77.3 & ---   &  78.4 &  75.2 & ---   & ---   \\
Fe\,{\sc ii} & 4656.98 & 2.89 & -3.55 & ---   & ---   & ---   & ---   & ---   &  60.5 & ---   & ---   & ---   & ---   & ---   & ---   \\
Fe\,{\sc ii} & 5414.08 & 3.22 & -3.68 &  28.1 &  36.2 &  34.5 &  36.6 &  25.5 &  37.2 &  41.7 &  37.7 &  37.7 &  37.3 &  33.4 &  35.2 \\
Fe\,{\sc ii} & 5991.38 & 3.15 & -3.56 &  35.9 & ---   &  45.8 &  45.9 &  34.7 & ---   &  48.5 & ---   & ---   &  42.5 &  44.1 &  43.8 \\
Fe\,{\sc ii} & 6084.11 & 3.20 & -3.81 &  25.3 &  30.6 &  31.2 &  31.9 &  23.4 &  35.7 &  34.7 &  35.1 &  35.7 &  33.0 &  34.0 & ---   \\
Fe\,{\sc ii} & 6149.26 & 3.89 & -2.72 &  35.6 &  49.1 &  44.6 &  45.6 &  34.5 &  51.3 &  51.1 &  48.2 &  49.4 &  43.5 &  46.0 &  46.1 \\  
\hline
\end{tabular}
\end{turn}
\end{center}
\end{table*}

\begin{table*}
\begin{center}
\caption{IC 4756 Ni and Zn equivalent widths \label{table:EW_list_4}}
\begin{turn}{90}
\begin{tabular}{lccccccccccccccc}
\hline
             &          &   &         & Her6  & Her35 & Her82 & Her87 & Her90 & Her144& Her170& Her176& Her228& Her249& Her296& Her397\\
             & $\lambda$&   &         & EW    & EW    & EW    & EW    & EW    & EW    & EW    & EW    & EW    & EW    & EW    & EW    \\
Element      & (\AA)    & EP&$\log gf$& (m\AA)& (m\AA)& (m\AA)& (m\AA)& (m\AA)& (m\AA)& (m\AA)& (m\AA)& (m\AA)& (m\AA)& (m\AA)& (m\AA)\\
\hline       
Ni\,{\sc i}  & 4831.17 & 3.61 & -0.32 & 108.5 & 102.9 &  99.3 &  95.4 & ---   & 109.9 &  93.4 &  91.9 &  97.5 &  96.4 &  94.3 &  96.9 \\
Ni\,{\sc i}  & 4857.39 & 3.74 & -0.83 &  77.7 &  70.2 &  64.4 &  69.1 & ---   &  74.0 &  65.8 &  59.4 &  72.3 &  70.2 &  69.3 &  66.4 \\
Ni\,{\sc i}  & 4866.26 & 3.54 & -0.21 & 111.9 & 113.9 &  99.9 & 102.1 & ---   & 111.3 &  99.9 &  90.2 & 111.8 & 108.5 & 100.3 & 106.4 \\
Ni\,{\sc i}  & 4904.41 & 3.54 & -0.25 & 117.5 & 107.7 & 102.4 & 102.6 & ---   & 113.7 &  94.0 &  92.8 & 105.3 &  99.9 &  94.7 & 105.4 \\
Ni\,{\sc i}  & 4953.20 & 3.74 & -0.58 &  93.9 &  82.9 &  80.1 &  78.5 & ---   &  84.3 &  74.6 &  73.6 &  81.9 &  79.1 &  73.7 &  73.2 \\
Ni\,{\sc i}  & 4998.22 & 3.61 & -0.69 &  86.0 &  83.6 &  74.0 &  77.3 & ---   &  90.0 &  76.1 &  75.5 &  85.9 &  81.0 &  77.0 &  79.2 \\
Ni\,{\sc i}  & 5082.34 & 3.66 & -0.54 &  94.1 &  87.9 &  81.1 &  85.5 & ---   &  88.9 &  82.0 &  77.2 &  89.3 &  83.5 &  82.5 &  80.6 \\
Ni\,{\sc i}  & 5084.09 & 3.68 & -0.07 & ---   & 111.4 &  98.9 & 111.1 & ---   & 116.2 &  98.3 &  97.2 & 106.4 & 104.0 &  99.1 & 106.8 \\
Ni\,{\sc i}  & 5088.53 & 3.85 & -1.06 &  58.9 &  53.8 &  50.2 &  54.1 & ---   &  51.1 &  49.0 &  46.7 &  50.4 &  54.4 &  51.5 &  50.9 \\
Ni\,{\sc i}  & 5088.95 & 3.68 & -1.29 &  52.6 &  59.3 &  50.1 &  50.6 & ---   &  49.4 &  49.0 &  49.7 &  51.2 &  49.1 &  53.2 &  43.1 \\    
Ni\,{\sc i}  & 5094.41 & 3.83 & -1.11 &  55.2 &  50.6 &  45.7 &  55.9 & ---   &  50.9 &  51.4 &  49.7 &  49.0 &  53.3 &  44.2 &  49.0 \\    
Ni\,{\sc i}  & 5102.96 & 1.68 & -2.67 & 104.4 &  95.3 &  82.9 &  84.8 & ---   &  98.4 &  86.7 &  81.3 &  93.6 &  85.0 &  84.1 &  83.2 \\
Ni\,{\sc i}  & 5115.39 & 3.83 & -0.13 & 108.9 &  99.8 &  95.1 &  89.8 & ---   & 107.8 &  89.0 &  91.3 &  99.3 &  94.3 &  95.2 &  96.8 \\ 
Ni\,{\sc i}  & 5392.33 & 4.15 & -1.31 &  27.0 &  23.9 &  27.0 & ---   &  31.6 &  24.0 & ---   & ---   & ---   &  25.2 &  25.4 &  26.3 \\
Ni\,{\sc i}  & 5468.10 & 3.85 & -1.66 &  30.1 &  22.4 &  21.3 &  27.6 & ---   &  29.9 &  29.0 &  28.1 & ---   &  22.3 & ---   &  23.7 \\  
Ni\,{\sc i}  & 5578.71 & 1.68 & -2.56 & 110.2 & 100.5 &  92.7 &  97.0 & 115.0 & 110.0 &  94.6 &  83.5 & 100.4 &  99.6 &  92.8 &  92.1 \\
Ni\,{\sc i}  & 5587.85 & 1.94 & -2.45 & 105.9 &  96.5 &  88.8 &  89.6 & 108.6 &  95.9 &  84.7 &  83.2 &  87.9 &  88.3 &  85.0 &  84.5 \\
Ni\,{\sc i}  & 5593.73 & 3.90 & -0.77 &  70.8 &  69.5 &  62.0 &  67.3 &  70.4 &  70.3 &  59.7 &  59.3 &  63.3 &  63.7 &  60.5 &  59.1 \\
Ni\,{\sc i}  & 5748.35 & 1.68 & -3.24 &  79.1 &  71.5 &  61.0 &  71.1 &  90.8 &  68.4 &  64.9 &  65.3 &  62.3 &  63.4 &  63.7 &  68.3 \\
Ni\,{\sc i}  & 5846.99 & 1.68 & -3.45 &  76.4 &  62.4 &  54.2 &  60.6 & ---   &  59.0 &  60.2 &  59.4 &  60.0 &  60.3 &  54.8 &  51.1 \\  
Ni\,{\sc i}  & 5996.73 & 4.24 & -1.00 &  41.1 &  43.1 &  36.0 &  36.7 &  45.0 &  35.0 &  35.0 &  34.4 &  36.7 &  38.9 &  38.0 &  31.5 \\    
Ni\,{\sc i}  & 6007.31 & 1.68 & -3.40 &  80.6 &  59.2 &  52.8 &  55.6 &  76.5 &  63.1 &  53.2 &  53.4 &  63.8 &  57.2 &  60.2 &  55.7 \\
Ni\,{\sc i}  & 6086.28 & 4.27 & -0.45 &  64.9 &  66.7 &  64.1 &  64.0 &  66.7 &  62.2 &  55.7 &  57.0 &  68.0 &  64.1 &  56.0 &  61.2 \\ 
Ni\,{\sc i}  & 6108.11 & 1.68 & -2.44 & 124.4 & 112.2 &  99.5 & 108.9 & 134.1 & 111.7 & 105.8 &  98.1 & 112.9 & 103.2 &  95.5 & 109.3 \\    
Ni\,{\sc i}  & 6111.07 & 4.09 & -0.82 &  57.8 &  60.4 &  53.8 &  54.7 &  63.0 &  58.5 &  53.7 &  56.5 &  57.9 &  50.0 &  55.7 &  50.7 \\  
Ni\,{\sc i}  & 6128.96 & 1.68 & -3.36 &  75.6 &  71.1 &  58.5 &  62.9 &  85.9 &  66.4 &  62.3 &  63.6 &  66.3 &  62.9 &  61.9 &  58.4 \\
Ni\,{\sc i}  & 6130.13 & 4.27 & -0.89 &  44.4 &  41.5 &  40.2 &  42.0 &  41.2 &  41.2 &  41.0 &  36.8 &  37.7 &  40.0 &  40.2 &  40.7 \\
Ni\,{\sc i}  & 6175.36 & 4.09 & -0.50 &  75.6 &  70.6 &  69.5 &  70.8 &  77.7 &  77.4 &  62.6 &  68.5 &  73.6 &  69.7 &  65.9 &  70.9 \\
Ni\,{\sc i}  & 6176.81 & 4.09 & -0.26 &  93.5 &  91.5 &  84.3 &  81.3 &  94.4 &  86.0 &  80.5 &  75.0 &  77.0 &  83.6 &  74.1 &  85.2 \\  
Ni\,{\sc i}  & 6177.24 & 1.83 & -3.55 &  57.8 &  42.8 &  46.0 &  39.7 &  59.9 &  43.7 &  44.1 &  40.7 &  47.6 &  39.6 &  43.5 &  35.0 \\
Ni\,{\sc i}  & 6186.71 & 4.11 & -0.88 &  58.7 &  55.7 &  45.3 &  46.3 &  58.0 &  49.3 &  53.2 &  48.3 &  55.8 &  51.0 &  50.6 &  47.4 \\  
Ni\,{\sc i}  & 6204.60 & 4.09 & -1.10 &  50.3 &  42.0 &  35.8 &  38.8 &  41.8 &  38.3 &  42.3 &  40.2 &  35.9 &  37.0 &  43.7 &  34.4 \\
Ni\,{\sc i}  & 6223.98 & 4.11 & -0.91 &  48.8 &  54.7 &  48.5 &  49.1 & ---   &  47.0 &  42.6 &  43.6 &  50.6 &  48.8 &  46.7 &  42.0 \\  
Ni\,{\sc i}  & 6314.65 & 1.94 & -1.99 & 131.2 & 119.1 & 111.5 & 113.2 & 134.5 & 123.1 & 107.9 & 101.3 & 119.2 & 118.5 & 106.0 & 112.7 \\   
Ni\,{\sc i}  & 6322.16 & 4.15 & -1.21 &  37.3 &  38.7 &  25.0 &  33.5 &  37.1 &  31.1 &  29.2 &  29.7 &  28.5 &  30.5 &  32.2 &  28.3 \\
Ni\,{\sc i}  & 6327.59 & 1.68 & -3.06 &  96.6 &  84.1 &  80.5 &  79.7 & 100.9 &  84.2 &  78.5 &  74.5 &  76.1 &  79.8 &  74.0 &  78.0 \\
Ni\,{\sc i}  & 6378.25 & 4.15 & -0.81 &  54.4 &  58.9 &  53.9 &  50.8 &  59.4 &  60.1 &  45.2 &  49.4 &  57.4 &  48.4 &  50.7 &  52.9 \\
Ni\,{\sc i}  & 6482.80 & 1.94 & -2.76 &  99.4 &  89.3 &  79.2 &  81.4 & 105.3 &  91.8 &  79.5 &  70.4 &  83.9 &  80.7 &  73.1 &  77.4 \\
Ni\,{\sc i}  & 6598.59 & 4.24 & -0.90 &  50.5 &  48.4 &  44.2 &  42.8 &  54.0 &  47.1 &  41.8 &  37.0 &  41.8 &  48.1 &  44.4 &  40.2 \\
Ni\,{\sc i}  & 6635.12 & 4.42 & -0.72 &  41.2 &  43.4 &  43.9 &  41.8 &  52.2 &  47.3 &  42.6 &  42.3 &  46.4 &  37.5 &  42.9 &  42.8 \\
Ni\,{\sc i}  & 6643.63 & 1.68 & -1.92 & 168.9 & 142.4 & 131.7 & 141.4 & 164.3 & 152.7 & 126.8 & 127.1 & 142.1 & 140.7 & 130.2 & 135.9 \\
Ni\,{\sc i}  & 6767.77 & 1.83 & -2.10 & 142.3 & 125.2 & 114.1 & 117.3 & 149.8 & 127.0 & 117.4 & 112.9 & 124.7 & 117.2 & 111.4 & 122.7 \\
Ni\,{\sc i}  & 6772.31 & 3.66 & -0.93 &  80.6 &  77.6 &  70.4 &  73.8 &  83.9 &  77.3 &  69.1 &  64.4 &  78.5 &  70.9 &  74.3 &  78.0 \\ 
\\
Zn\,{\sc i}  & 4810.53 & 4.08 & -0.26 &  87.4 &  90.7 &  85.4 &  83.8 & ---   &  96.6 &  83.2 &  76.6 &  89.6 &  84.8 &  84.8 &  87.9 \\
Zn\,{\sc i}  & 6362.34 & 5.80 &  0.12 &  21.2 &  29.2 &  26.9 &  28.4 &  21.3 &  27.3 &  28.8 &  29.2 &  29.7 &  28.8 &  28.8 &  27.1 \\
\hline
\end{tabular}
\end{turn}
\end{center}
\end{table*}


\label{lastpage}


\begin{thebibliography}{99}

\bibitem[\protect\citeauthoryear{Adibekyan et al.}{2012}]{adi12} Adibekyan V. Zh. et al., 2012, \aap, 545, A32
\bibitem[\protect\citeauthoryear{Alonso, Arribas \& Mart{\'{\i}}nez-Roger}{Alonso et al.}{1999}]{alo99} Alonso A., Arribas S., Mart{\'{\i}}nex-Roger C., 1999, \aaps, 140, 261
\bibitem[\protect\citeauthoryear{Andrievsky et al.}{2008}]{and08} Andrievsky S.~M., Spite M., Korotin S.~A., Spite F., Bonifacio P., Cayrel R., Hill V., Fran\c{c}ois P., 2008, \aap, 481, 481
\bibitem[\protect\citeauthoryear{Bensby, Feltzing \& Lundstr{\"o}m}{Bensby et al.}{2003}]{ben03} Bensby T., Feltzing S., Lundstr{\"o}m I., 2003, \aap, 410, 527
\bibitem[\protect\citeauthoryear{Bisterzo et al.}{2010}]{bis10} Bisterzo S., Gallino R., Straniero O., Cristallo S., K{\"a}ppeler F., 2010, \mnras, 404, 1529
\bibitem[\protect\citeauthoryear{Bland-Hawthorn, Krumholz \& Freeman}{Blank-Hawthorn et al.}{2010}]{bla10} Bland-Hawthorn J., Krumholz M.~R., Freeman K.~C., 2010, \apj, 713, 166
\bibitem[\protect\citeauthoryear{Bonifacio et al.}{2009}]{bon09} Bonifacio P. et al., 2009, \aap, 501, 519
\bibitem[\protect\citeauthoryear{Bovy, Rix \& Hogg}{Bovy et al.}{2012a}]{bov12a} Bovy J., Rix H.~W., Hogg D.~W., 2012a, \apj, 751, 131
\bibitem[\protect\citeauthoryear{Bovy et al.}{2012b}]{bov12b} Bovy J., Rix H.~W., Liu C., Hogg D.~W., Beers T.~C., Lee Y.~S., 2012b, \apj, 753, 148
\bibitem[\protect\citeauthoryear{Bovy et al.}{2012c}]{bov12c} Bovy J., Rix H.~W., Hogg D.~W., Beers T.~C., Lee Y.~S., Zhang L., 2012c, \apj, 755, 115
\bibitem[\protect\citeauthoryear{Bubar \& King}{2010}]{bub10} Bubar E.~J., King J.~R., 2010, \aj, 140, 293
\bibitem[\protect\citeauthoryear{Busso, Gallino \& Wasserburg}{Busso et al.}{1999}]{bus99} Busso M., Gallino R., Wasserburg G.~J., 1999, \araa, 37, 239
\bibitem[\protect\citeauthoryear{Busso et al.}{2001}]{bus01} Busso M., Gallino R., Lambert D.~L., Travaglio C., Smith V.~V., 2001, \apj, 557, 802
\bibitem[\protect\citeauthoryear{Carrera \& Pancino}{2011}]{car11} Carrera R., Pancino E., 2011, \aap, 535, A30
\bibitem[\protect\citeauthoryear{Castelli \& Kurucz}{1994}]{cas94} Castelli F., Kurucz R.~L., 1994, \aap, 281, 817
\bibitem[\protect\citeauthoryear{Chieffi \& Limongi}{2002}]{chi02} Chieffi A., Limongi M., 2002, \apj, 577, 281
\bibitem[\protect\citeauthoryear{Cristallo et al.}{2009}]{cri09} Cristallo S., Straniero O., Gallino R., Piersanti L., Dom{\'i}nguez I., Lederer M.~T., 2009, \apj, 696, 797
\bibitem[\protect\citeauthoryear{D'Orazi \& Randich}{2009}]{dor09} D'Orazi V., Randich S., 2009, \aap, 501, 553
\bibitem[\protect\citeauthoryear{De Silva et al.}{2006}]{siv06} De Silva G.~M., Sneden C., Paulson D.~B., Asplund M., Bland-Hawthorn J., Bessell M.~S., Freeman K.~C., 2006, \aj, 131, 455
\bibitem[\protect\citeauthoryear{De Silva et al.}{2007a}]{siv07a} De Silva G.~M., Freeman K.~C., Asplund M., Bland-Hawthorn J., Bessell M.~S.,Collet, R., 2007a, \aj, 133, 1161
\bibitem[\protect\citeauthoryear{De Silva et al.}{2007b}]{siv07b} De Silva G.~M., Freeman K.~C., Bland-Hawthorn J., Asplund M., Bessell M.~S., 2007b, \aj, 133, 694
\bibitem[\protect\citeauthoryear{De Silva et al.}{2009}]{siv09} De Silva G.~M., Gibson, B. K., Lattanzio J., Asplund M., 2009, \aap, 500, l25
\bibitem[\protect\citeauthoryear{De Silva et al.}{2011}]{siv11} De Silva G.~M., Freeman K.~C., Bland-Hawthorn J., Asplund M., Williams M., Holmberg J., 2011, \mnras, 415, 563
\bibitem[\protect\citeauthoryear{Freeman \& Bland-Hawthorn}{2002}]{free02} Freeman K.~C., Bland-Hawthorn J., 2002, \araa, 40, 487
\bibitem[\protect\citeauthoryear{Friel, Jacobson \& Pilachowski}{Friel et al.}{2010}]{fri10} Friel E.~D., Jacobson H.~R., Pilachowski C.~A., 2010, \aj, 139, 1942
\bibitem[\protect\citeauthoryear{Gilroy}{1989}]{gil89} Gilroy K.~K., 1989, \apj, 347, 835
\bibitem[\protect\citeauthoryear{Grevesse \& Sauval}{1998}]{gre98} Grevesse N., Sauval A.~J., 1998, in Fr{\"o}hlich C., Huber M.~C.~E., Solanki S.~K., von Steiger R., eds, Proc. ISSI Workshop, Solar Composition and Its Evolution-- From Core to Corona. Kluwer, Dordrecht, p. 161
\bibitem[\protect\citeauthoryear{Herwig}{2005}]{her05} Herwig F., 2005, \araa, 43, 435
\bibitem[\protect\citeauthoryear{Herzog, Sanders \& Seggewiss}{Herzog et al.}{1975}]{her75} Herzog A.~D., Sanders W.~L., Seggewiss W., 1975, \aaps, 19, 211
\bibitem[\protect\citeauthoryear{Jacobson, Friel \& Pilachowski}{Jacobson et al.}{2007}]{jac07} Jacobson H.~R., Friel E.~D., Pilachowski C.~A., 2007, \aj, 134, 1216
\bibitem[\protect\citeauthoryear{Jacobson, Friel \& Pilachowski}{Jacobson et al.}{2011}]{jac11} Jacobson H.~R., Friel E.~D., Pilachowski C.~A., 2011, \aj, 141, 58
\bibitem[\protect\citeauthoryear{K{\"a}ppeler et al.}{2011}]{kap11} K{\"a}ppeler F., Gallino R., Bisterzo S., Aoki W., 2011, \rmp, 83, 157
\bibitem[\protect\citeauthoryear{Karakas}{2010}]{kar10} Karakas A. I., 2010, \mnras, 403, 1413
\bibitem[\protect\citeauthoryear{Kobayashi et al.}{2006}]{kob06} Kobayashi C., Umeda H., Nomoto K., Tominaga N., Ohkubo T., 2006, \apj, 653, 1145
\bibitem[\protect\citeauthoryear{Lind et al.}{2011}]{lind}  Lind K., Asplund M., Barklem P.~S., Belyaev A.~K.,2011, \aap, 528, 103
\bibitem[\protect\citeauthoryear{Luck}{1994}]{luc94} Luck R.~E., 1994, \apjs, 91, 309
\bibitem[\protect\citeauthoryear{Mashonkina et al.}{2007}]{mas07} Mashokina L.~I., Vinogradova A.~B., Ptitsyn D.~A., Khokhlova V.~S., Chernetsova T. A., 2007, \arep, 51, 903
\bibitem[\protect\citeauthoryear{Pace et al.}{2010}]{pac10} Pace G., Danziger J., Carraro G., Melendez J., Frac{\c c}ois P., Matteucci F., Santos N.~C., 2010, \aap, 515, A28
\bibitem[\protect\citeauthoryear{Pancino et al.}{2010}]{pan10} Pancino E., Carrera R., Rossetti E., Gallart C., 2010, \aap, 511, A56
\bibitem[\protect\citeauthoryear{Reddy et al.}{2003}]{red03} Reddy B.~E., Tomkin J., Lambert D.~L., Allende Prieto C., 2003, \mnras, 340, 304
\bibitem[\protect\citeauthoryear{Reddy, Lambert \& Allende Prieto}{Reddy et al.}{2006}]{red06} Reddy B.~E., Lambert D.~L., Allende Prieto C., 2006, \mnras, 367, 1329
\bibitem[\protect\citeauthoryear{Roederer \& Sneden}{2011}]{roe11} Roederer I.~U., Sneden C., 2011, \aj, 142, 22
\bibitem[\protect\citeauthoryear{Salaris, Weiss \& Percival}{Salaris et al.}{2004}]{sal04} Salaris M., Weiss A., Percival S.~M., 2004, \aap, 414, 163
\bibitem[\protect\citeauthoryear{Santos et al.}{2009}]{san09} Santos N.~C., Lovis C., Pace G., Melendez J., Naef D., 2009, \aap, 493, 309 
\bibitem[\protect\citeauthoryear{Schmidt}{1978}]{sch78} Schmidt E.~G., 1978, \pasp, 90, 157
\bibitem[\protect\citeauthoryear{Schuler et al.}{2003}]{sch03} Schuler S.~C., King J.~R., Fischer D.~A., Soderblom D.~R., Jones B.~F., 2003, \aj, 125, 2085
\bibitem[\protect\citeauthoryear{Shi, Gehren \& Zhao}{Shi et al.}{2004}]{shi} Shi J.R., Gehren T., Zhao G., 2004, \aap, 423, 683
\bibitem[\protect\citeauthoryear{Smiljanic et al.}{2009}]{smi09} Smiljanic R., Gauderon R., North P., Barbuy B., Charbonnel C., Mowlavi N., 2009, \aap, 502, 267
\bibitem[\protect\citeauthoryear{Smith}{1983}]{smi83} Smith G.~H., 1983, \pasp, 95, 296
\bibitem[\protect\citeauthoryear{Thogersen, Friel \& Fallon}{Thogersen et al.}{1993}]{tho93} Thogersen E.~N., Friel E.~D., Fallon B.~V., 1993, \pasp, 105, 1253
\bibitem[\protect\citeauthoryear{Ting et al.}{2012}]{ting12} Ting Y.~S., Freeman K.~C., Kobayashi C., De Silva G.~M., Bland-Hawthorn J., 2012, \mnras, 421, 1231
\bibitem[\protect\citeauthoryear{Umeda \& Nomoto}{2002}]{ume02} Umeda H., Nomoto K., 2002, \apj, 565, 385
\bibitem[\protect\citeauthoryear{Umeda \& Nomoto}{2005}]{ume05} Umeda H., Nomoto K., 2005, \apj, 619, 427
\bibitem[\protect\citeauthoryear{Vassiliadis \& Wood}{1993}]{vas93} Vassiliadis E., Wood P. R., 1993, \apj, 413, 641
\bibitem[\protect\citeauthoryear{Woosley \& Weaver}{1995}]{woo95} Woosley S.~E., Weaver T.~A., 1995, \apjs, 101, 181
\bibitem[\protect\citeauthoryear{Yong, Carney \& Teixera de Almeida}{Yong et al.}{2005}]{yon05} Yong D., Carney B.~W., Teixera de Almeida M.~L., 2005, \aj, 130, 597


\end{thebibliography}
\end{document}